\documentclass[aps,pre,reprint,showpacs,floatfix]{revtex4-1}

\usepackage{lmodern,amsmath,amssymb,mathtools}
\usepackage{graphicx}
\usepackage[utf8]{inputenc}
\usepackage{siunitx}
\usepackage[colorlinks=true,pdfusetitle]{hyperref}
\usepackage{todonotes}

\newcommand{\ie}{\emph{i.e.}}
\newcommand{\eg}{\emph{e.g.}}
\newcommand{\etal}{\emph{et al.}}
\newcommand{\df}{\mathrm{d}}
\newcommand{\LL}{{\cal{L}}}
\newcommand{\vv}[1]{\boldsymbol{#1}}
\newcommand{\te}[1]{\text{#1}}
\DeclarePairedDelimiter{\abs}{\lvert}{\rvert}
\DeclarePairedDelimiter{\avg}{\langle}{\rangle}
\DeclareMathOperator{\erfc}{erfc}

\begin{document}

\title{Analytical mesoscale modeling of aeolian sand transport}

\date{\today}
\author{Marc L\"ammel}
\author{Klaus Kroy}

\email[Corresponding author. \\ Electronic address:]{klaus.kroy@uni-leipzig.de} 
\affiliation{Institut f\"ur Theoretische Physik, Universit\"at Leipzig, Postfach 100920, 04009 Leipzig, Germany}

\begin{abstract}
  The mesoscale structure of aeolian sand transport determines a variety of natural phenomena studied in planetary and Earth science.
  We analyze it theoretically beyond the mean-field level, based on the grain-scale transport kinetics and splash statistics.
  A coarse-grained analytical model is proposed and verified by numerical simulations resolving individual grain trajectories.
  The predicted height-resolved sand flux and other important characteristics of the aeolian transport layer agree remarkably well with a comprehensive compilation of field and wind tunnel data, suggesting that the model robustly captures the essential mesoscale physics.
  By comparing the saturation length with field data for the minimum sand-dune size, we can reconcile conflicting previous models for this most enigmatic emergent aeolian scale and elucidate the importance of intermittent turbulent wind fluctuations for field measurements.
\end{abstract}

\maketitle
\section{Introduction}
\label{sec:introduction}
Aeolian sand transport is the process of erratic grain hopping occasionally observed on a windy day at the beach.
It remains perplexing how the wide variety of distinctive aeolian sand patterns, from tiny ripples to huge dunes, emerges from such seemingly chaotic dynamics.
The current knowledge about the grain-scale structure of aeolian transport largely rests on laboratory and field experiments~\cite{Rasmussen2015}.
Attempts to derive coarse-grained mathematical models that can rationalize the observations started with Bagnold's seminal work in the 1930s~\cite{Bagnold1941} and are still the subject of ongoing research~\cite{Duran2011,Kok2012,Paehtz2013,Barchyn2014,Jenkins2014,Berzi2016,Mayaud2017,Martin2017,Berzi2017}, for good conceptual and practical reasons.
With average grain trajectories exceeding the sub-millimeter grain scale by orders of magnitude, aeolian transport is a typical mesoscale phenomenon that should be amenable to such coarse-grained modeling.
Moreover, despite growing computational resources, faithful grain-scale simulations remain forbiddingly expensive, so that also numerical approaches cannot avoid fairly drastic idealizations~\cite{Anderson1988,Kok2009,Duran2012,Carneiro2013,Paehtz2015}.
And even a perfect simulation of aeolian transport would \emph{per se}, without a theory, not make the emergence of the various mesoscales and ensuing sand structures less mysterious.

A radically coarse-grained mean-field model~\cite{Sauermann2001} that maps the whole mobilized grain population onto a single effective grain trajectory has been successful in explaining desert dune formation~\cite{Kroy2002,Kroy2002a,Andreotti2002a,Duran2010a}.
Such mean-field approaches roughly account for the more energetic saltating grains but fail to resolve the heterogeneity of the transport layer, which contains a majority of grains that only perform very small hops and do not eject other grains from the bed~\cite{Ungar1987}.
Their less spectacular transport mode, conventionally referred to as bedload, reptation, or creep, is however thought to be largely responsible for ripple and megaripple formation~\cite{Bagnold1941,Anderson1987,Hoyle1999,Yizhaq2004a,Manukyan2009,Duran2014}, which therefore eludes the mean-field approaches.
Also ecologically important processes, such as dust emission, rock abrasion, and vegetation invasion, are sensitive to the detailed mesostructure of aeolian transport~\cite{Yizhaq2007,Kok2012}.
Finally, an improved theoretical model of the aeolian transport layer could help to infer more information about extraterrestrial conditions from the limited data obtained by remote-sensing~\cite{Sullivan2005,Bridges2012,Ayoub2014,Burr2015}.

In the following, we propose a way to transcend the mean-field approximation and to account, with good precision, not only for the mean transport characteristics but also for the substantial heterogeneity and the fluctuations within the transport layer and in the turbulent wind.
Similar as in earlier contributions~\cite{Ungar1987,Anderson1988,Werner1990,Anderson1991,Kok2009a}, the ballistic kinetics of an ensemble of wind-blown grains is coupled to the local wind strength and to the dissipative collisions with the sand bed.
However, as one crucial novel key ingredient, we utilize a recently proposed analytical model for the grain--bed collisions~\cite{Laemmel2017} that was extensively tested against grain-scale computer simulations and laboratory experiments.
It enables us to develop a neat analytical description for the whole distribution of grain trajectories (Sec.~\ref{sec:reduced-model}), from which explicit formulas for various height-resolved observables, like the grain concentration and flux, readily follow (Sec.~\ref{sec:results}).
In line with results from earlier numerical work~\cite{Kok2010,Kok2012}, the wind strength is found to affect the transport-layer physics only weakly, as corroborated by various wind-tunnel studies~\cite{Dong2003,Creyssels2009,Ho2011,Ho2014,Rasmussen2015}.
It enters primarily via ``global'', height-integrated quantities and the total height of the transport layer, which are amenable to conventional mean-field transport models~\cite{Owen1964,Sauermann2001,Sorensen2004,Duran2006a,Laemmel2012,Paehtz2012}.
To validate key ingredients of the analytical modeling, we performed dedicated coarse-grained computer simulations that also explicitly resolve the broad distribution of grain trajectories.
This feature turns out to be crucial for a proper analysis of a large amount of field and laboratory data, as shown in Sec.~\ref{sec:literature-data}, where the model predictions are thoroughly tested against literature data.
The field and wind-tunnel data suggest that the mesoscale structure of the aeolian transport layer is well captured by the analytical model.
On this basis, we can make a strong case for a proportionality of the minimum sand-dune size reported in field measurements to the so-called saturation length for flux transients in heterogeneous wind, which is however somewhat masked by the renormalizing effect of intermittent turbulent wind-strength fluctuations that we explicitly include in the data analysis.
Throughout the main text, we emphasize conceptual aspects and defer technical details to six appendices.

\section{Analytical model}
\label{sec:reduced-model}
The crux of our approach is to condense the height-resolved ``mesoscale'' information about the heterogeneities in the stationary aeolian transport layer into a simple analytical form of the hop-height distribution:
\begin{equation}
  \label{eq:hopheight-distribution}
  P_H(h) \propto T(h) h^{-\nu}  \te e^{-h/H} \,.
\end{equation}
Here, \(T(h)\) denotes the hop time for the trajectory of height \(h\), \(H\) is the (wind-dependent) characteristic height of the transport layer, and the power-law exponent \(\nu\) quantifies the energy distribution of ejected bed grains.
The combination \(h^{-\nu} \te e^{-h/H}\) refers to the probability to find a trajectory of height \(h\), weighting it with the flight time \(T(h)\) then yields the probability that a randomly selected grain in the transport layer follows such a trajectory.
Approximating
\begin{equation}
  \label{eq:hop-time}
  T(h) \approx 2\sqrt{2h/g} \,,
\end{equation}
by the free flight time (neglecting vertical drag), with \(g\) being the gravitational acceleration, yields
\begin{equation}
  \label{eq:drag-free-hopheight-distribution}
  P_H(h) = [H \Gamma(3/2-\nu)]^{-1} (h/H)^{1/2-\nu} \te e^{-h/H} \,.
\end{equation} 
The Euler gamma function \(\Gamma\) arises upon normalization.

We now provide physical arguments for the proposed form of \(P_H(h)\).
Near the bed, the transport statistics is dominated by the large number of ejected grains, whose energies are log-normally distributed and very small compared to the energy of the saltating grain ejecting them~\cite{Ho2012,Laemmel2017}.
Their hop-height distribution is therefore well captured by the asymptotic power-law relation \(P_H(h\ll H) \propto h^{-\nu}\) with
\begin{equation}
  \label{eq:hopheight-nu}
  \nu = 1-2/\log(4)^2+1/\log(2)
  \approx 1.4 \,,
\end{equation}
as shown in App.~\ref{sec:splash-predictions}.
In contrast, large hop heights (\(h>H\)) are only reached by the few particles that survive many bed collisions, and are therefore expected to die out exponentially.
More precisely, the fixed rebound probability \(P_\te{reb} \approx P_\te{rep}(h \gg a) \approx 0.86\) for trajectories much higher than the grain diameter \(a\) (typically a few \SI{100}{\micro m}) and the mean ejection height \(\overline{h_\te{ej}} \approx 11 a\), both computed from our splash model~\cite{Laemmel2017} in App.~\ref{sec:splash-predictions}, yield \(P_H(h \gg \overline{h_\te{ej}}) \propto T(h) \te e^{-h/H}\) with
\begin{equation}
  \label{eq:char-hopheight}
  H \approx -\overline{h_\te{ej}}/\ln P_\te{reb} \approx 72a \,.
\end{equation}
We also note that an alternative explanation for the form of \(P_H(h)\) in Eq.~\eqref{eq:drag-free-hopheight-distribution} invokes an analogy with the barometer formula for the concentration of a thermalized gas~\cite{Jenkins2010}.

\begin{figure}
  \centering
  \includegraphics{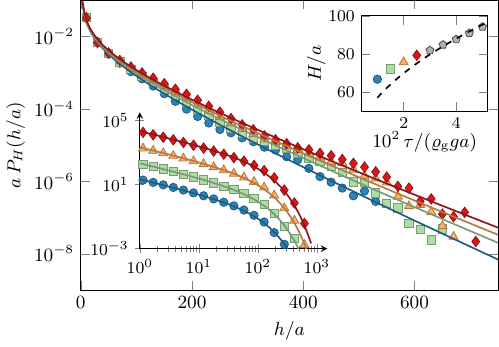}
  \caption{Coarse-grained numerical simulations (symbols) of the hop-height distribution \(P_H(h)\) compared to the analytical model, Eq.~\eqref{eq:drag-free-hopheight-distribution} (solid lines).
    The simulation computes the statistics of the ejected bed grains from the analytical model of Ref.~\cite{Laemmel2017}, which yields a power-law distribution for low hop heights with exponent \(\nu = 1.4\) (log-log representation in the left inset).
    \emph{Right inset}: trajectories reaching beyond the characteristic height \(H\) of the transport layer die out exponentially.
    While \(H\) was, for each data point, used as a free fit parameter to match the simulation data, its \(\tau\)-dependence is well reproduced by the more refined calculation in Sec.~\ref{sec:wind-depend-transp} (dashed line).}
  \label{fig:salsim-hopheightdistr}
\end{figure}

To further back up Eq.~\eqref{eq:drag-free-hopheight-distribution}, we performed coarse-grained computer simulations, as described in App.~\ref{sec:simulations}.
The comparison between analytical theory and simulations in Fig.~\ref{fig:salsim-hopheightdistr} shows excellent agreement if the transport layer height \(H\) is employed as a free fit parameter that slightly increases with increasing wind shear stress \(\tau\).
The right inset of Fig.~\ref{fig:salsim-hopheightdistr} compares the extracted \(H(\tau)\) with the refined calculation presented in Sec.~\ref{sec:wind-depend-transp}.
But even the simpler version of the model, which disregards this weak wind-dependence and fixes \(H\) to its characteristic value according to Eq.~\eqref{eq:char-hopheight}, will suffice for many practical purposes.

The model proposed in Eq.~\eqref{eq:drag-free-hopheight-distribution} has manifold consequences and applications.
First, note that the hop-height distribution \(P_H(h)\) together with the purely ballistic conditional probability \(P(z|h)\) for grains on a trajectory of height \(h\) to be found at height \(z\) make up the joint probability \(P_H(z,h) = P(z|h)P_H(h)\) for grains to be at height \(z\) on a trajectory of height \(h\).
Using Eq.~\eqref{eq:hop-time} for the free flight time and writing \(\Theta\) for the Heaviside step function, \(P(z|h)\te d z = [2/T(h)]\te d t\) yields
\begin{equation}
  \label{eq:single-trajectory-concentration}
  2 h P(z|h) =
  \Theta(h-z)/\sqrt{1 - z/h} \,.
\end{equation}
The ensuing joint probability \(P_H(z,h)\) then provides us with the height-dependent profiles
\begin{subequations}
  \label{eq:profiles-def}
  \begin{align}
    \label{eq:concentration-profile-def}
    \varrho_H (z) &= \rho_H \int \!\! \te d h \, P_H(z,h) \,,\\
    \label{eq:hor-flux-profile-def}
    j_H(z) &= \rho_H \int \!\! \te d h \, P_H(z,h) v_x(z,h) \,, \\
    \label{eq:ver-flux-profile-def}
    \phi_{H}(z,\ell) &= \frac{\rho_H}{2} \!\!\!\! \int \limits_{l(h)>\ell} \!\!\!\! \!\! \te d h P_H(z,h) \abs{v_z(z,h)} \,, \\
    \label{eq:grain-stress-profile-def}
    \tau_{\te g, H}(z) &= \rho_H \int_z^\infty \!\! \!\!\!\! \df \tilde z \int \!\! \te d h \,  P_H(\tilde z,h) f_x(\tilde z, h) \,, 
  \end{align}
\end{subequations}
of the mass concentration, the horizontal and vertical fluxes, and the stress contribution of the sand grains, respectively.
Taken together, these functions allow for a comprehensive characterization of the height-resolved mesoscale structure of the transport layer, so that the underlying model assumptions can be tested with unprecedented scrutiny.
For brevity, we have introduced some additional notation in Eqs.~\eqref{eq:profiles-def}, namely the horizontal and vertical components \(v_x(z,h)\), \(v_z(z,h)\) of the mean velocity, and the horizontal component \(f_x(z,h)\) of the force (per grain mass) contributed by a trajectory of height \(h\) to the grain-borne shear stress~\cite{Ungar1987}.
The common overall scale factor \(\rho_H=\int\!\te d z \varrho_H(z)\) is the height-integrated mass concentration in the saltation layer (units mass/area).
The factor 1/2 in Eq.~\eqref{eq:ver-flux-profile-def} arises because each trajectory contributes twice to the grain density \(P_H(z,h)\), namely during ascent and descent, whereas it contributes only once to the local vertical flux \(\phi_H(z,\ell)\), either when ascending or descending.
Also note that we restricted the ensemble of trajectories in Eq.~\eqref{eq:ver-flux-profile-def} to those with total length \(l(h)>\ell\), in order to make contact with experiments that use horizontal sand traps to measure the vertical flux \(\phi_H(z=0,\ell)\) through the sand bed at a downwind distance \(\ell\) from the end of the bed~\cite{Namikas2003,Ho2014,Rasmussen2015}.

The general form
\begin{equation}
  \label{eq:general-integral-structure}
  \int \!\! \df h\, K(z,h) P_H(z,h) = \int_z^\infty \!\!\! \!\!\!\! \df h \, K(z,h) \frac{h^{-1/2-\nu} \te e^{-h/H}}{2\sqrt{1-z/h}}
\end{equation}
of the integrals in Eqs.~\eqref{eq:profiles-def} already allows some general conclusions to be drawn as to how much the distribution and shape of the grain trajectories matters for a given mesoscale observable.
Namely, unless one is specifically interested in the conditions very near the ground (\(z/H \ll 1\)), the singularity of the denominator for \(h\to z\) gives large weight to the corresponding value of the rest of the integrand.
The latter can then be taken out of the integral and only the characteristic height \(H\) of the transport layer matters, if \(K(z,h)\) does not accidentally cancel the square-root singularity.
This actually happens in Eqs.~\eqref{eq:ver-flux-profile-def}, \eqref{eq:grain-stress-profile-def}; the vertical flux \(\phi_H(z,\ell)\) and the grain-borne stress \(\tau_\te{g,H}(z)\) are thus sensitive to the precise \(h\)-dependent shape of the short trajectories with \(h\ll H\), while mass concentration and horizontal flux are not.

To explicitly evaluate the expressions in Eqs.~\eqref{eq:profiles-def}, it is useful to make a few relatively uncritical simplifications concerning the shape and kinematics of the grain trajectories.
First, as in Eq.~\eqref{eq:hop-time}, we again neglect the vertical drag for the individual trajectory, using the free-flight estimate for the vertical velocity for ascent and descent.
The mean horizontal velocity then follows if the trajectories are approximated as parabolas of aspect ratio \(\epsilon(h) = h/l(h)\):
\begin{subequations}
  \label{eq:velocity-estimates}
  \begin{align}
      \label{eq:ver-velocity-vs-h}
 v_z^2(z,h) &\approx 2g(h-z) \\
    \label{eq:hor-velocity-vs-h}
    v_x(z,h)  \approx v_x(h) & \approx l(h)/T(h) \approx \sqrt{2gh}/[4\epsilon(h)] 
  \end{align}
\end{subequations}
Marked deviations only occur for the few very long trajectories with \(h \gg H\).
These ``flyers'' are exposed to the unscreened wind speed that increases approximately logarithmically with \(h\).
They therefore acquire exceptionally high forward speeds \(v_x(z,h)\) and strongly asymmetric trajectories, but  have little effect on typical observables, due to their rare occurrence.

Disregarding them for now, each hopping grain can be said to be accelerated by the wind drag during most of its flight time.
Since it passes each height \(z\) twice, its mean horizontal velocity \(v_x(z,h)\) should thus, to a fair approximation, be \(z\)-independent, as indeed confirmed by numerically solving the equations of motion of a representative hopping grain (see, \eg, Ref.~\cite{Jenkins2014}).
The forward speed during descent and ascent can then be written as \(v_x(h) \pm \Delta v_x(z,h)/2\), with the net wind-induced speed-up (to first order in \(z\)) given by
\begin{equation}
  \label{eq:ver-velocity-increase-vs-h}
  \Delta v_x(z,h) \approx \alpha v_x(h) (1 - z/h) \,.
\end{equation}
Here, the momentum loss incurred upon rebound is represented by a constant (\(h\)-independent) effective restitution coefficient \(\alpha = \Delta v_x(z=0,h)/v_x(h)\).
The velocity increment \(\Delta v_x(z,h)\) directly determines the force 
\begin{equation}
  \label{eq:grain-force}
  f_x(z,h)
  = -\frac{\df v_x(z,h)}{\df t}
  = -\abs{v_z(z,h)} \frac{\partial \Delta v_x(z,h)}{2\partial z}\,
\end{equation}
entering the grain-borne shear stress via Eq.~\eqref{eq:grain-stress-profile-def}.
With Eqs.~\eqref{eq:hor-velocity-vs-h}, \eqref{eq:ver-velocity-increase-vs-h}, the outer integral in Eq.~\eqref{eq:grain-stress-profile-def} thus reads
\begin{equation}
  \label{eq:integrated-grain-force}
  \int_z^\infty \!\!\!\!\!\!\! \df \tilde z P(\tilde z, h) f_x(\tilde z,h) 
  = P(0,h) v_z(0,h) \Delta v_x(z,h)/2\,.
\end{equation}

In the above expressions for the forward velocity and derived quantities, the aspect ratio \(\epsilon(h) = h/l(h)\) of the trajectories remains to be specified.
In cases where only the shape of the typical trajectory matters, it suffices to estimate \(\epsilon(H)\).
Neglecting its weak wind-speed dependence, we consider conditions near the transport threshold, where few particles are mobilized and the bare logarithmic wind-speed profile prevails.
Combining it with the prediction of our splash model for the typical rebound geometry of the hopping grains, we then find
\begin{equation}
  \label{eq:epsH}
  \epsilon(H) \approx 0.1 \,,
\end{equation}
in App.~\ref{sec:splash-predictions}.
While this estimate suffices for many practical purposes, the aspect ratios of the shortest and longest trajectories, corresponding to so-called reptating particles and flyers, respectively, deviate from \(\epsilon(H)\) according to 
\begin{equation}
  \label{eq:aspect-ratio-regimes}
  \frac{\epsilon(h)}{\epsilon(H)} \approx
  \begin{cases}
    (h/H)^{-1/2} & \qquad (h \ll H) \\
    1 & \qquad (h\simeq H) \\
    0 & \qquad (h\gg H)
  \end{cases}
\end{equation}
The three regimes, which are clearly discernible in the simulations (see Fig.~\ref{fig:salsim-aspectratio} of App.~\ref{sec:simulations}), can be interpreted in terms of three asymptotically dominant transport modes, namely almost vertically splashed grains near the ground, saltating grains that attain a limiting speed near the top of the transport layer, and a few flyers above it.
The power-law stretching for the shortest trajectories follows from the typical hop length of wind-blown ejected bed grains, as detailed in App.~\ref{sec:splash-predictions}.
With increasing height, the flight time \(T(h)\) approaches the typical response time (``drag time'') for relaxation to the stationary velocity, so that one might expect the aspect ratio to evolve roughly as \(h/[T(H)u(h)] \propto \sqrt h/\ln(h/z_0)\), using the logarithmic law of the wall for the wind speed \(u(h)\) at height \(h\) with surface roughness scale \(z_0\).
However, such an argument neglects the vertical drag that becomes increasingly important as \(h\) increases.
It limits the vertical velocity (essentially to the terminal velocity) and thus effectively caps the hop height of the fastest ejecta.
The highest trajectories, beyond an intermittent regime of approximately shape-invariant trajectories for \(h \simeq H\), will therefore ultimately become increasingly stretched.
According to Eq.~\eqref{eq:drag-free-hopheight-distribution}, these flyers are so rare that they have little impact on typical mesoscale observables, so that the first two regimes in Eq.~\eqref{eq:aspect-ratio-regimes} suffice to derive a wealth of accurate analytical predictions.
Nevertheless, an improved scheme that relaxes the shape invariance of the grain trajectories to also estimate the height dependence of flyer-sensitive observables for \(h>H\), is proposed below, in Sec.~\ref{sec:wind-depend-transp}.

With all the ingredients in place, we can finally estimate the average value of the characteristic transport layer \(H\) from a flux-balance argument.
Under stationary transport conditions, the vertical grain flux \(\phi_H(z=0,\ell=0)\) into the bed must compensate for the outgoing (rebounding and ejected) grains, yielding zero net erosion~\cite{Andreotti2004,Kok2010}.
With the above notation, this criterion reads
\begin{equation}
  \label{eq:vert-flux-balance}
  \int_a^\infty \!\!\!\!\!\! \te d h \left[ 1- P_\te{reb}(h) - N(h) \right] \abs{v_z(0,h)} P_H(0,h) = 0\,,
\end{equation}
where the grain size \(a\) is used as minimum hop height to regularize the otherwise unbounded integral.
Precise relations for the rebound probability \(P_\te{reb}\) and the number \(N\) of ejected grains per impact can be taken from our splash model~\cite{Laemmel2017}.
It provides them as functions of the impact energy, approximately \(mgh[1+1/(4\epsilon)^2]\), which in turn depends on the tangent of the impact angle, here rephrased as \(4\epsilon\), according to Eq.~\eqref{eq:velocity-estimates}.
Together with \(\epsilon = 0.1\) from Eq.~\eqref{eq:epsH}, the implicit equation~\eqref{eq:vert-flux-balance} for the average transport layer height is readily solved and yields \(H \approx 70 a\).
The quantitative agreement of this result with Eq.~\eqref{eq:char-hopheight}, which only relies on the rebound characteristics of the splash model \cite{Laemmel2017}, is no surprise, since it is dominated by the high-hopping and therefore frequently rebounding grains (rather than by the ejecta).
If desired, the neglected weak wind-strength dependence of \(H\) through \(N\) and \(\epsilon\) can be included in the analysis using the extended approach of Sec.~\ref{sec:wind-depend-transp}.

At this point, we would like to emphasize that all system parameters introduced above were computed from our splash model, which was independently calibrated in Ref.~\cite{Laemmel2017} by comparison with collision experiments using plastic beads.
However, to compare the various predictions derived in the following section with laboratory and field data in Sec.~\ref{sec:literature-data}, below, the characteristic height \(H\) is better used as a free fit parameter, to compensate for the  differences between the actual sand grains and said plastic beads, as well as for some poorly controlled environmental conditions such as humidity and temperature.
Our comparison with field data also reveals that the value of the power-law exponent \(\nu\) characterizing the splash efficiency should be adapted to the actual experimental setup, whereas no adjustment is required for the values of both the aspect ratio \(\epsilon(H) \approx 0.1\) of the typical trajectories and the effective restitution coefficient \(\alpha \approx 0.6\), computed in App.~\ref{sec:splash-predictions}.

\section{Results}
\label{sec:results}
\subsection{Height-resolved transport characteristics}
\label{sec:height-resolved-transport}
\begin{figure}
  \centering
  \includegraphics{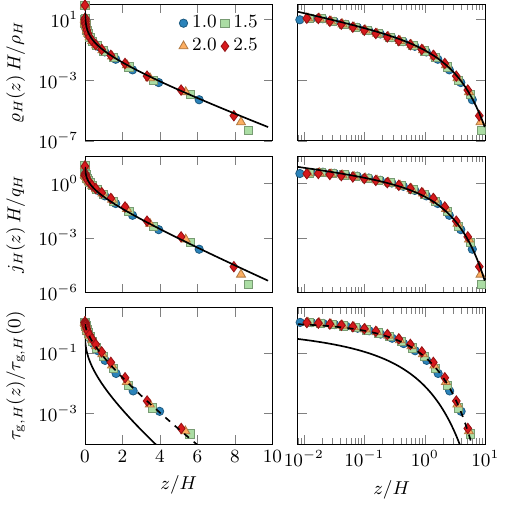}
  \caption{Simulation data (symbols) and analytical predictions obtained from Eqs.~\eqref{eq:drag-free-hopheight-distribution}--\eqref{eq:ver-velocity-increase-vs-h} (lines) for the height-resolved mass concentration \(\varrho_H(z)\), flux \(j_H(z)\), and grain-borne shear stress \(\tau_{\te g, H}(z)\).
    Normalizing the data by the height-integrated concentration \(\rho_H\) and flux \(q_H\), the grain stress at the ground \(\tau_{\te g, H}(0)\), and the characteristic transport-layer height \(H\), a good collapse is achieved for various wind shear stresses \(\tau\) (values for \(10^2 \cdot \tau/(\varrho_\te g g a)\) in the legend), as theoretically predicted.
    The solid lines correspond to the model approximation of shape-invariant trajectories with \(\epsilon(h) = \epsilon(H)\) independent of \(h\); the dashed lines represent Eq.~\eqref{eq:grain-stress}, which invokes \(\epsilon(h) \propto h^{-1/2}\), as appropriate for the majority of (low) trajectories, which collectively carry most of the momentum and therefore dominate the stress (see the inset of Fig.~\ref{fig:salsim-aspectratio}).}
  \label{fig:salsim-profiles}
\end{figure}

We now come back to the height-resolved observables introduced in Eqs.~\eqref{eq:profiles-def}.
Explicit analytical expressions for the height-resolved grain concentration and flux are given in App.~\ref{sec:shape-inv-traj}.
They take the asymptotic forms
\begin{equation}
  \label{eq:concentration-profile}
  \frac{\varrho_H(z)}{\rho_H/H} \propto
  \begin{cases}
     (z/H)^{1/2-\nu}   & \quad (z \ll H)\\
    (z/H)^{-\nu}  \te e^{-z/H}  & \quad (z \gg H)\\
\end{cases}
\end{equation}
and
\begin{equation}
  \label{eq:flux-height-profile}
  \frac{j_H(z)}{q_H/H} \propto
  \begin{cases}
    (z/H)^{1-\nu}  & \quad (z \ll H)\\
    (z/H)^{1/2-\nu}  \te e^{-z/H}  & \quad (z \gg H)\,,
  \end{cases}
\end{equation}
with the height-integrated flux \(q_H \propto \rho_H v_x(H)\).
Here and for the plots in Fig.~\ref{fig:salsim-profiles}, we inserted Eq.~\eqref{eq:hor-velocity-vs-h} for \(v_x(h,z)\) and assumed shape-invariant trajectories with \(\epsilon(h) = \epsilon(H)\).
The very good agreement, for all \(z\), between model and numerical data in the upper and middle panels of Fig.~\ref{fig:salsim-profiles} supports our simplifying assumptions.

Similarly, the height-resolved grain-borne shear stress \(\tau_{\te g, H}(z)\) is estimated by inserting Eqs.~\eqref{eq:velocity-estimates}--\eqref{eq:ver-velocity-increase-vs-h} into Eq.~\eqref{eq:grain-stress-profile-def}.
As anticipated above, one here needs to account for the shape of the short trajectories \(\epsilon(h \ll H) \propto (h/H)^{1/2}\), corresponding to the the first regime of Eq.~\eqref{eq:aspect-ratio-regimes},
\begin{equation}
  \label{eq:grain-stress}
  \frac{\tau_{\te g,H}(z)}{\tau_{\te g,H}(0)}
  \approx \frac{\Gamma(2-\nu,z/H) - (z/H) \Gamma(1-\nu,z/H)}{\Gamma(2-\nu)} \,,
\end{equation}
where \(\tau_{\te g,H}(0) = \alpha g \rho_H/[8 \epsilon(H)]\).
Although this result rests on the assumption \(h \ll H\), it provides a very good estimate for arbitrary values of \(z\), because the exponential decay of the gamma functions for \(z \gg H\) dominates over any polynomial \(h\)-dependence originating from \(\epsilon(h)\).
Physically speaking, the condition \(h \ll H\) comprises the majority of the grain trajectories, which account for almost the whole grain-borne shear stress.
The quality of this prediction is illustrated in the lower two panels of Fig.~\ref{fig:salsim-profiles}.

Given any general relation \(h(l)\) between trajectory height and length, the vertical grain flux at arbitrary height \(z\) reads
\begin{equation}
  \label{eq:vert-flux-profile}
  \frac{\phi_H(z,\ell)}{q_H/L}
  = \frac{ \Gamma\left(1-\nu, \max \!\left[ \left\{z/H, h(\ell)/H \right\} \right] \right)}{\Gamma(3/2-\nu)} \,,
\end{equation}
with \(L = H/\epsilon(H)\) being the length of trajectories of height \(H\).
For \(h \gg H\), the incomplete gamma function decays exponentially, so that \(\phi_H(z,\ell) \propto [h(\ell)/H]^{-\nu} \te e^{-\ell(h)/H}\) (tacitly assuming \(z<h\)).
Taking the derivative of Eq.~\eqref{eq:vert-flux-profile} with respect to \(\ell\) at \(z=0\) yields the hop-length distribution
\begin{equation}
  \label{eq:hop-length-distribution}
  -\frac{\partial}{\partial \ell}\frac{\phi_H(0,\ell)}{\phi_H(0,0)} \propto \frac{1}{H} \frac{\df h(\ell)}{\df \ell} [h(\ell)/H]^{-\nu} \te e^ {-h(\ell)/H}\,.
\end{equation}
It is usually measured in experiments with sand traps~\cite{Ho2014,Rasmussen2015} or in simulations by counting every trajectory once, independent of its length and hop time.
It can also be directly inferred from the hop-height distribution in Eq.~\eqref{eq:hopheight-distribution}, after dropping the hop time \(T(h)\).
Inserting the asymptotic scaling \(\epsilon(h \ll H) \propto (h/H)^{-1/2}\) from Eq.~\eqref{eq:aspect-ratio-regimes} into Eq.~\eqref{eq:hop-length-distribution}, we find \(h(\ell)/H \approx (\ell/L)^{2/3}\).
From this, the power-law exponent of the hop-length distribution follows as \(1/3 + 2 \nu/3 \approx 1.27\).
A slightly smaller exponent \(1.2\) was indeed observed by Dur\'an \etal~\cite{Duran2014} in full-fledged grain-scale computer simulations of the saltation process.

The model predictions in Eqs.~\eqref{eq:concentration-profile}--\eqref{eq:hop-length-distribution} can now be used to derive explicit expressions for further height-dependent transport properties.
Combining Eqs.~\eqref{eq:concentration-profile} and \eqref{eq:flux-height-profile}, for instance, immediately yields the height dependence of the mean grain velocity
\begin{equation}
  \label{eq:velocity-height-profile}  
    V_H (z) = j_H(z)/\varrho_H(z) \propto v_x(H) \sqrt{z/H}\,.
\end{equation}
Note however, that the division makes the result sensitive to the precise functional form of the singular growth and the small tails of \(j_H(z)\) and \(\rho_H(z)\) for very small and large \(z\), respectively.
Therefore, it can only be trusted for intermediate heights \(z\), as suggested by its square-root growth consistent with the regime of approximately shape-invariant saltation trajectories (the next section shows how this limitation can be overcome).
In this regime, our model moreover provides an explicit prediction for the full height-resolved velocity distribution
\begin{equation}
  \label{eq:velocity-disitribution}
  P_H(v_x \vert z) = P_H[z,h(z,v_x)] \abs*{\partial_{v_x} h(z,v_x)} \rho_H /\varrho_H(z) \,,
\end{equation}
which follows from the variable transformation \(h = h(z,v_x)\) together with the general relation \(P_H(z,v_x) = P_H(v_x \vert z) \varrho_H(z)/\rho_H\) between joint and conditional distribution.
Exploiting the shape invariance of the trajectories that dominate the statistics at intermediate \(z\), we have \(v_x(h)^2 = (h/H) v_x(H)^2 \) and thus
\begin{equation}
  \label{eq:horiz-velocity-distribution}
  P_H(v_x|z) =
  \frac{\rho_H  [v_x/v_x(H)]^{-2\nu} \te e^{-v_x^2/v_x(H)^2}}{\Gamma(3/2-\nu) H \varrho_H(z) \sqrt{v_x^2-v_x(H)^2z/H}} \,.
\end{equation}
A similar calculation yields the distribution of the vertical grain velocity
\begin{equation}
  \label{eq:vert-velocity-distribution}
  P_H(v_z|z) = \frac{\rho_H [z/H + v_z^2/(2gH)]^{-\nu} \te e^{z/H + v_z^2/(2gH)}}{\Gamma(3/2-\nu) H \varrho_H(z) v_z(0,H)} \,.
\end{equation}

\begin{figure}
  \centering
  \includegraphics{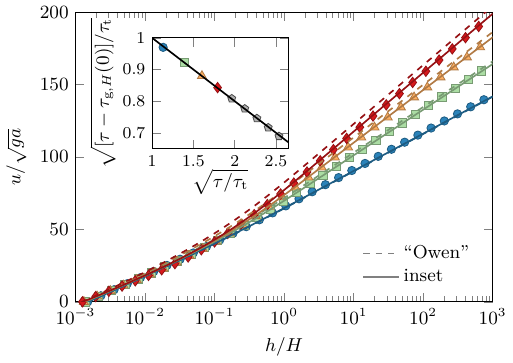}
  \caption{Height-resolved wind speed from numerical simulations (symbols) and our model, Eq.~\eqref{eq:wind-height-profile}, with (dashed lines) and without (solid lines) invoking the simplifying Owen hypothesis that the air-borne shear stress at the bed is screened precisely to the threshold value \(\tau_\te t\) for grain entrainment~\cite{Owen1964}.}
  \label{fig:salsim-windspeedprofile}
\end{figure}

Our model also allows for a reliable analytical estimate of the wind-speed profile \(u(z)\) within and above the transport layer, by considering the feedback of the grain-borne momentum on the logarithmic law of the wall~\cite{Owen1964}.
Inserting Eq.~\eqref{eq:grain-stress} into the modified Prandtl turbulence closure \(\tau - \tau_{\te g, H} (z) = \varrho_\te a \kappa^2 z^2 [\df u_H(z)/\df z]^2\), where \(\varrho_\te a\) is the air density and \(\kappa \approx 0.4\) the von K\'arm\'an constant, and approximating~\cite{Laemmel2012}
\begin{equation}
  \label{eq:grain-stress-profile-approx}
  \sqrt{1- \tau_{\te g, H}(z)/\tau} \approx 1 - \left[ 1-\sqrt{1- \tau_{\te g, H}(0)/\tau} \right]  \frac{\tau_{\te g, H}(z)}{\tau_{\te g, H}(0)} \,
\end{equation}
we obtain the height-dependent wind velocity
\begin{equation}
  \label{eq:wind-height-profile}
  \begin{split}
    u_H(z) &= \frac{u_\ast}{\kappa} \ln(z/z_0) - \frac{u_\ast - u_{\ast \te t}}{\kappa}
    \omega(z_0/H,z/H) \\
    &\sim
    \begin{dcases}
      \frac{u_{\ast\te t}}{\kappa} \ln(z/z_0)  & (z \ll H) \\
      \frac{u_{\ast\te t}}{\kappa} \ln(z_\te f/z_0) + \frac{u_\ast}{\kappa} \ln(z/z_\te f) & (z \gg H)
    \end{dcases}
  \end{split}
\end{equation}
with \(\omega(\zeta_0,\zeta_1) \equiv \int_{\zeta_0}^{\zeta_1} \df \zeta \, \tau_{\te g,H}(\zeta H)/[\zeta \tau_{\te g,H}(0)]\), the roughness height \(z_0 \approx a/10\) of the (quiescent) send bed, and the shear velocity \(u_\ast \equiv \sqrt{\tau/\varrho_\te a}\).
Following Ref.~\cite{Owen1964}, we here approximated the air shear stress at the ground  as a wind-strength independent constant \(\tau-\tau_{\te g,H}(0) \approx \tau_\te t \equiv \varrho_\te a u_{\ast \te t}^2\), which is justified near the transport threshold.
As illustrated in Fig.~\ref{fig:salsim-windspeedprofile}, this (``zeroth-order'') approximation is already in good qualitative agreement with the numerical simulations, while they can be perfectly matched when small first-order corrections in \(\sqrt{\tau/\tau_\te t}-1\) ~\cite{Duran2006,Kok2012} are taken into account.

The asymptotic relations in Eq.~\eqref{eq:wind-height-profile} follow from \(\tau_{\te g, H}(z\ll H) \approx \tau_{\te g, H}(0)\) and \(\tau_{\te g, H}(z\gg H) \approx 0\), according to Eq.~\eqref{eq:grain-stress}.
They correspond to the two limits of the screened logarithmic wind-speed profile near the ground~\cite{Laemmel2012} and the ``shifted'' logarithmic profile above the transport layer, which can also be rewritten as \(u_H(z) \sim (u_\ast/\kappa) \ln(z/z_0^\te{eff})\) with the wind-strength dependent effective roughness length \(z_0^\te{eff} = z_0 (z_\te f/z_0)^{1-u_{\ast\te t}/u_\ast}\).
The crossover height
\begin{equation}
  \label{eq:focus-height}
  z_\te f \equiv z_0 \te e^{\omega(z_0/H,\infty)} \approx H \te e^{\Gamma'(2-\nu)/\Gamma(2-\nu)-1} \approx 0.1 H
\end{equation}
between the two is comparable to the average hop height
\begin{equation}
  \label{eq:mean-hop-height}
   \int \! \df h \, h P_H(h) = (3/2-\nu) H \approx 0.1 H
\end{equation}
of a randomly chosen grain in a transport layer of height \(H\), and also to the median height \(z_\te m \approx 0.2 H\) of its flux profile, indirectly defined through
\(q_H/2 = \int_0^{z_\te m} \df z \, j_H(z)\) (App.~\ref{sec:two-spec}).
Recalling that \(H\) depends only weakly on the wind strength (Fig.~\ref{fig:salsim-hopheightdistr}), we thus interpret \(z_\te f\) as the so-called ``focus height''~\cite{Bagnold1941,Andreotti2004,Duran2006}.

Altogether, our results corroborate theoretically that the mesoscale structure of aeolian transport is characterized by a very dense layer of hopping particles at the ground, which is only a few grain diameters high (\(\approx 0.1 H \approx 7 a\)).
The layer is predominantly populated by so-called reptating grains, which have hardly gained momentum from the wind.
This is a direct consequence of the power-law decay of the hop-height distribution in Eq.~\eqref{eq:drag-free-hopheight-distribution}, which in turn is strictly tied to the splash statistics of the bed collisions~\cite{Laemmel2017}.
In contrast the grains on typical saltation trajectories are accelerated by the wind, giving them an essentially length-independent shape.
This symmetry is only broken by the longest trajectories of the flyers that reach heights \(h\gg H\) far beyond the characteristic height \(H\) of the transport layer.
It is this drastic change between the transport modes of the hopping particles as a function of their jump height, which limits most severely the application of mean-field models based on a single effective trajectory and has motivated the development of two-species approaches~\cite{Andreotti2004,Laemmel2012}.

\subsection{Broken shape invariance: height-dependent grain velocity}
\label{sec:broken-shape-invariance}
In the previous section, we noted that the power-law increase of the average grain velocity \(V_x(z)\) with height \(z\), as obtained in Eq.~\eqref{eq:velocity-height-profile}, is restricted to intermediate \(z\), where the shape-invariant trajectories contribute most.
As we show now, this limitation can be overcome by an alternative approach that relates \(V_x(z)\) to the height-dependent wind speed \(u_H(z)\) from Eq.~\eqref{eq:wind-height-profile}.
We first note that the height-dependent average grain speed \(V_H(z) = j_H(z)/\varrho_H(z)\) is well approximated by the velocity \(v_x(z,h=z)\equiv v_x(z)\) at the apex of a representative trajectory of overall height \(z\).
Technically, this is seen by applying Laplace's method to the integrals for \(j_H(z)\) and \(\varrho_H(z)\) (see App.~\ref{sec:shape-inv-traj}).
The actual form of \(v_x(h)\) can then be estimated from the velocity gain \(\Delta v_x(z=0,h) = \int_0^{T(h)} \df t\, \dot v_x(t)\) of a grain during one hop of height \(h\):
\begin{equation}
  \label{eq:velocity-increase-force-balance}
  \Delta v_x(0,h)
  = \frac{\sqrt{2gh}}{v_\infty^2} \int \!\! \df \tilde z \, P_H(\tilde z | h) [u_H(\tilde z) - v_x(\tilde z,h)]^2
\end{equation}
Here, the conditional probability \(P_H(\tilde z | h) \propto 1/v_z(\tilde z, h)\) results from the transformation \(\df \tilde z = v_z \df t \) of the integration variable, as in Eq.~\eqref{eq:single-trajectory-concentration}, and we used a single integral as a shorthand representation of both the ascending and descending part of the trajectory.
The remaining terms on the right-hand side represent the drag force on the grain---see Eq.~\eqref{eq:sim:eom} of App.~\ref{sec:simulations}.
We exploited the stretched shape of high trajectories, which implies \(v_z \ll v_x , u\), to drop the vertical velocity components, and expressed the drag coefficient in terms of the turbulent settling velocity \(v_\infty\).
For typical dune sand with an average grain diameter \(a\approx \SI{200}{\micro m}\), \(v_\infty \approx 27 \sqrt{ga}\)~\cite{Jimenez2003}.
The integral in Eq.~\eqref{eq:velocity-increase-force-balance} can roughly be approximated using the fact that \(P_H(\tilde z | h)\) diverges for \(\tilde z \to h\); again, only the value \(v_x(h,h)\) of the grain speed at the apex (\(z=h\)) matters, while the shape of the whole trajectory is found to be irrelevant (and can thus not be determined within this approach).
Inserting \(P_H(\tilde z | h) \approx 2\delta(\tilde z - h)\Theta(h-\tilde z)\) in Eq.~\eqref{eq:velocity-increase-force-balance} and making contact with Eqs.~\eqref{eq:ver-velocity-increase-vs-h} and \eqref{eq:hor-velocity-vs-h}, which yield \(\Delta v_x(0,h) \propto \sqrt h\), we eventually obtain
\begin{equation} 
  \label{eq:velocity-force-balance}
  V_H(z) \approx u_H(z) - \sqrt{\alpha/[4\epsilon(z)]} v_\infty \qquad (z \gg H)\,.
\end{equation}
This is of the same form as the force-balance estimate obtained by Sauermann \etal~\cite{Sauermann2001} for a representative grain that is exposed to turbulent drag and an additional (constant) bed friction.
With the typical values for \(\alpha \approx 0.6\) and \(\epsilon(z \approx H) \approx 0.1\) that follow from our splash model, the ensuing prediction for the height-dependent grain velocity, Eq.~\eqref{eq:velocity-force-balance}, is found to be in very good agreement with our simulations for the whole range \(z>z_\te{f}\), above the focus height (data not shown).

\subsection{Wind-dependent transport layer height}
\label{sec:wind-depend-transp}
In Eq.~\eqref{eq:char-hopheight} of Sec.~\ref{sec:reduced-model}, we combined the splash-model predictions for the rebound probability and the average ejection height to estimate the overall value of the characteristic height \(H \approx 72 a\) of the transport layer.
This estimate was supported by our simulations, shown in the inset of Fig.~\ref{fig:salsim-hopheightdistr}, which, however, also revealed that \(H\) increases weakly the wind strength.
And, based on the accurate prediction for the grain velocity \(V_x(z)\) in Eq.~\eqref{eq:velocity-force-balance}, we indeed expect such a wind-strength dependence of \(H\).
To calculate it quantitatively, it seems tempting to follow the relation between \(h\) and \(v_x\) given in Eq.~\eqref{eq:hor-velocity-vs-h}, namely \(H \approx 8 [\epsilon(H_0) V_{H_0}(H_0)]^2/g \), with \(H_0 = 72 a\) and \(\epsilon(H_0)\approx 0.1\) being the wind-independent predictions of Eqs.~\eqref{eq:char-hopheight} and \eqref{eq:epsH}, respectively.
Our simulations show that the wind dependence of \(\epsilon(H)\) partly compensates for the one of \(V_H(H)\), such that their product \(\epsilon(H) V_H(H)\) increases relatively weakly with the wind strength, but we could not obtain a good analytical estimate for the combined effect.
Instead, a similar approach for the characteristic trajectory length,
\begin{equation}
  \label{eq:wind-depend-charact-trajectory}
  L \approx \tilde T(H_0) V_{H_0}(H_0) \quad \te{and} \quad
  H \approx \epsilon(H_0) L \,,
\end{equation}
turned out to work well if one allows for a slight phenomenological adjustment \(\tilde T(H_0) = 0.7 T(H_0)\) of the hop time compared to Eq.~\eqref{eq:hop-time}, in order to match the theory quantitatively to our simulation data.
Note that Eq.~\eqref{eq:wind-depend-charact-trajectory} leaves the characteristic aspect ratio \(H/L = \epsilon(H_0)\) independent of the wind strength, but the individual trajectories' shape invariance is broken, as expressed by the (yet unknown) dependence \(\epsilon(h)\).

Alternatively, the wind-strength dependence of \(H\) may be taken from one of the mean-field transport models mentioned in Sec.~\ref{sec:introduction}.
In App.~\ref{sec:two-spec}, this is illustrated for the two-species model of Ref.~\cite{Laemmel2012}, where the median height \(\approx 0.2 H\) of the transport layer---calculated after Eq.~\eqref{eq:mean-hop-height}---is used to feed its prediction into our present discussion.

\subsection{Flux transients}
\label{sec:flux-transients}
Finally, we want to address what is probably the most enigmatic and most debated mesoscale property of aeolian sand transport, the so-called saturation length \(\ell_\te{sat}\).
This central notion was originally introduced by Sauermann \etal~\cite{Sauermann2001} to quantify how the aeolian sand transport adapts to changes in the wind over uneven topographies.
Flux transients on the upwind slope of a sand dune and in the downwind wake region were discussed as two pertinent instances giving rise to quite diverse numerical values and parameter dependencies for
\(\ell_\te{sat}\).
Later work has further elaborated on this point~\cite{Andreotti2004,Claudin2006,Andreotti2010,Paehtz2013,Paehtz2015}.
As an emergent mesoscale concept, \(\ell_\te{sat}\) is thus intrinsically context-dependent, and attempts to promote narrower definitions of the saturation length (such as the distance needed for a hopping grain to be accelerated to the fluid velocity~\cite{Claudin2006} or the distance over which the sand flux saturates at the entrance of a sand bed~\cite{Andreotti2010}) seem counterproductive.

Arguably the most interesting saturation transients are those near the crest of a small dune, due to moderate local changes in the wind speed (rather than sand coverage).
They are responsible for the emergence of the relevant mesoscale \(\ell_\te{sat}^\te{dune}\) with respect to which aeolian dunes may be considered large or small, and which gives rise to a minimum dune length \(\LL_\te{min}\)~\cite{Kroy2002,Kroy2002a,Andreotti2002a,Kroy2005,Parteli2007a,Fourriere2010}.
In Ref.~\cite{Sauermann2001}, this particular length was predicted to decay with increasing wind strength \(\tau\), whereas later studies argued that it might either be independent of \(\tau\)~\cite{Andreotti2002,Andreotti2010} or even grow monotonically in \(\tau\)~\cite{Paehtz2013,Paehtz2015}.
To resolve the apparent conflict between these diverse proposals, we identify the saturation length in our simulations with the response length to a small wind-strength increment (to mimic the effect of the speed-up along the back of a dune within our stationary transport model).

\begin{figure}
  \centering
  \includegraphics{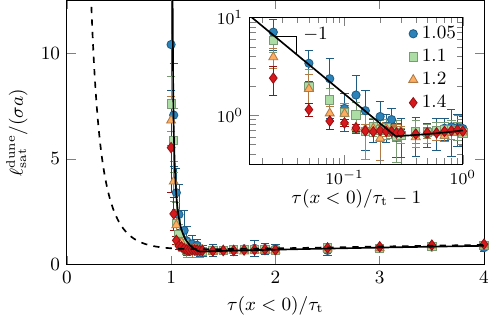}
  \caption{``Bare'' and ``dressed'' wind-strength dependent saturation lengths (linear and logarithmic axes scaling).
    Simulation data for the bare saturation length \(\ell_\te{sat}^\te{dune}\) (symbols) were obtained by fitting a linear relaxation process with decay length \(\ell_\te{sat}^\te{dune}\) to the response of the height-integrated flux \(q(x)\) along the wind direction to a small step increase of the shear stress \(\tau(x)\) at \(x=0\) (see legend for relative step height).
    Its power-law decay in \(\tau\) (inset) close to the transport threshold \(\tau_\te t = 0.01 \varrho_\te g g a\) gives way to a weak growth at larger \(\tau\), in good accord with Eqs.~\eqref{eq:wind-depend-charact-trajectory}, \eqref{eq:saturation-length}, with a numerical factor \(\ell_\te{sat}^\te{dune}/L = 2\) (solid lines).
    Under realistic field conditions, the sharp singularity at \(\tau = \tau_\te t\) is smeared out by intermittent wind-speed fluctuations, giving rise to the apparent ``dressed'' saturation length given by the right-hand side of Eq.~\eqref{eq:lsat-fluct-avg} (dashed line).}
  \label{fig:salsim-saturation-length}
\end{figure}

Due to the scale separation between the minimum dune size \(\LL_\te{min} \approx 10^5 a\) (see Fig.~\ref{fig:min-dune-size}) and the characteristic hop length \(L = H/\epsilon(H) \approx 7.2 \cdot 10^2 a\), as predicted by Eqs.~\eqref{eq:char-hopheight},~\eqref{eq:epsH}, the resulting flux gradients should not be sensitive to the heterogeneities inside the transport layer.
As a consequence, \(\ell_\te{sat}^\te{dune}\) can only depend on the overall transport characteristics and is thus expected to scale linearly in \(L\).
Indeed, our numerical data in Fig.~\ref{fig:salsim-saturation-length} are consistent with
\begin{equation}
  \label{eq:saturation-length}
  \ell_\te{sat}^\te{dune}/L \propto 1+\frac{\tau_\te{ta}-\tau}{\tau-\tau_\te t} \Theta(\tau_\te{ta}-\tau)  \,,
\end{equation}
and conforms with the expectation \(\ell_\te{sat}^\te{dune} \simeq L\) above the direct aerodynamic entrainment threshold (for \(\tau > \tau_\te {ta}\)).
Physically, mobile grains are then abundant and their acceleration to the stationary speed limits the adaptation to an increase in wind strength.
Notice that \(L\) itself slightly increases with the wind speed, according to Eq.~\eqref{eq:wind-depend-charact-trajectory}---see also Figs.~\ref{fig:salsim-hopheightdistr} and \ref{fig:vert-flux}.
If the wind speed falls below the direct entrainment threshold \(\tau_\te {ta}\) and approaches the transport threshold \(\tau_\te{t}\), grains need not only be accelerated, but an increasing amount of energy and momentum has to be supplied to lift grains from the ground against gravity.
The adaptation of the flux to an increase in wind speed is then limited by the number of grains  gradually mobilized in successive bed collisions.
The lift force exerted by the wind on the bed still facilitates the splash by effectively reducing the heavy mass of the grains, which can be understood as a precursor of direct aerodynamic entrainment~\cite{Anderson1991a}, as argued in Eq.~\eqref{eq:splash:drag-effective-mass} of App.~\ref{sec:splash-model}.
Accordingly, the net erosion rate scales with the excess shear stress \(\tau-\tau_\te t\), which vanishes at the transport threshold \(\tau_\te t\), causing a singular slowing-down of the adaptation of the flux to wind heterogeneities.
This manifests itself in the divergence of the saturation length \(\ell_\te{sat}^\te{dune}(\tau)\) at \( \tau_\te t\)~\cite{Sauermann2001}.
A similar decomposition of the physics behind the saturation length according to two rate-limiting processes was previously proposed by Andreotti \etal~\cite{Andreotti2010} to interpret wind-tunnel measurements.

It is important to realize that a direct application of Eq.~\eqref{eq:saturation-length}  to field data, is problematic, though.
Field data that do not conform with the intuitively expected scaling \(\LL_\te{min} \propto \ell_\te{sat}^\te{dune}\) for the minimum dune length \(\LL_\te{min}\) do not necessarily indicate a failure of the theory.
The reason is that the sharp peak of \(\ell_\te{sat}^\te{dune}\) at \(\tau=\tau_\te t\), with relative width \((\tau_\te{ta}-\tau_\te t)/\tau_\te t \approx 0.27\), is not resolved by typical field measurements, which inevitably average over some intermittent wind fluctuations.
The importance of a given wind strength for the observed structure formation is determined by the erosion or deposition caused, which are proportional to the flux.
We therefore propose that structural field data should be interpreted as flux-weighted averages of the corresponding ``bare'' theoretical predictions calculated for a fixed wind strength.
In particular, the minimum dune length observed in the field should not scale in the bare saturation length but in its ``dressed'' version, according to
\begin{equation}
  \label{eq:lsat-fluct-avg}
  \LL_\te{min} \propto 
  \avg{\phi\ell_\te{sat}^\te{dune} }_{\avg{\tau}} / \avg{\phi}_{\avg{\tau}} \,.
\end{equation}
The averages
\(\avg{\dots}_{\avg{\tau}}\) is understood to extend over a range of wind-strength fluctuations around the measured average shear stress \(\avg{\tau}\).
The bare vertical flux
\begin{equation}
\phi \equiv \phi_H(0,0) \propto q_H/L \propto (\tau-\tau_\te t)\Theta(\tau-\tau_\te t)/L
\end{equation}
vanishes at \(\tau <\tau_\te{t}\), thereby effectively truncating the wind-strength distribution.
The consequences of this are illustrated by the dashed line in Fig.~\ref{fig:salsim-saturation-length} assuming a realistic Weibull distribution of variance \(0.05\avg{u}^2\) (\(u \propto u_\ast \propto \sqrt{\tau}\)) for the wind-speed fluctuations~\cite{Pfeifer2012}.
For \(\tau \gg \tau_\te t\), the weakly wind-strength dependent bare saturation length \(\ell_\te{sat}^\te{dune}\) is hardly affected at all by the averaging: bare and dressed saturation length are indistinguishable in the plot.
In contrast, close to the transport threshold \(\tau_\te{t}\), the saturation length gets strongly renormalized by fluctuations.
The flux-averaged or dressed saturation length as a function of the average shear stress \(\avg \tau\)---the right-hand side of Eq.~\eqref{eq:lsat-fluct-avg}---has a strongly smeared-out shape compared to the bare prediction.

We incidentally find the dressed wind-strength dependence to be closely reminiscent of the form \(\ell_\te{sat}^\te{dune} \propto L/(\tau/\tau_\te t - 1)\) originally proposed by Sauermann \etal~\cite{Sauermann2001}.
This observation supports the interpretation of the measured (dressed) saturation length as an emergent hydrodynamic length scale and vindicates the use of Sauermann's formula in past analytical and numerical studies that are at the core of our current understanding of the physics of sand dunes~\cite{Parteli2014}.
Similar but potentially more drastic renormalizations may be expected in applications to extraterrestrial dunes, in which the gap between the threshold shear stresses \(\tau_\te t\) and \(\tau_\te{ta}\) can be much larger than on Earth~\cite{Claudin2006,Kok2010b}.

\section{Comparison with literature data}
\label{sec:literature-data}
After having established the accuracy of our analytical model by comparison with our simulation results, we now want to test it against a compilation of literature data.

\begin{figure}
  \centering
  \includegraphics{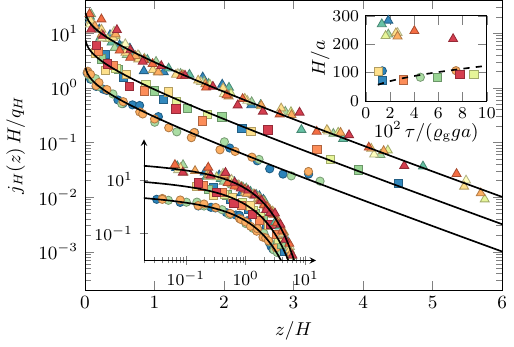}
  \caption{Theory and literature data for the height-resolved horizontal grain flux \(j_H(z)\)  (linear/logarithmic height axis; data extracted from independent sources vertically shifted, for better visibility).
    The simple analytical model with shape-invariant grain trajectories, Eqs.~\eqref{eq:drag-free-hopheight-distribution}--\eqref{eq:ver-velocity-vs-h} (solid lines), is fitted to wind-tunnel measurements by Rasmussen and Mikkelsen~\cite{Rasmussen1998} (dots), Rasmussen and S{\o}rensen~\cite{Rasmussen2008} (squares), and field data by Namikas~\cite{Namikas2003} (triangles) for various wind strengths \(\tau\), using \(\nu\) and \(H\) as free global fit parameters.
    This yields \(\nu=0.94\) and the data for \(H\) shown in the right inset with the improved model prediction from Eq.~\eqref{eq:wind-depend-charact-trajectory} (dashed line).
    While laboratory data and theory agree well, field measurements  (triangles) consistently find higher trajectories.}
  \label{fig:flux-profiles}
\end{figure}

Experimental data for the height-resolved horizontal flux \(j_H(z)\) is usually approximated, with fair accuracy, by an exponential profile with a mean height on the order of a few centimeters that is almost independent of the wind strength~\cite{Rasmussen2015}.
This resonates well with our Eq.~\eqref{eq:flux-height-profile}.
However, there is apparently no clear consensus about possible deviations from the exponential form due to a possible deficit~\cite{Dong2006,Creyssels2009} or excess~\cite{Rasmussen1998,Nishimura2000,Namikas2003,Rasmussen2008,Cheng2015,Rasmussen2015} of grains near the ground; probably because of difficulties in determining the exact number of mobile grains in the dense lower transport region.
In particular, the particle tracking and laser scattering methods used in Refs.~\cite{Dong2006,Creyssels2009} seem prone to missing some of the mobile grains close to the bed (the ejected ``reptating'' grains), so that the flux in this region is most likely underestimated.
In contrast, sand trap measurements in Refs.~\cite{Rasmussen1998,Namikas2003,Rasmussen2008} reflect the pronounced near-bed excess contribution to \(j_H(z)\) that we expect from our model as a direct consequence of the splash statistics.
They are indeed in remarkable agreement with the prediction of Eqs.~\eqref{eq:drag-free-hopheight-distribution}--\eqref{eq:ver-velocity-vs-h} if \(\nu\) and \(H\) are treated as free fit parameters, as illustrated in Fig.~\ref{fig:flux-profiles}.
The relatively small value deduced for the best-fitting power-law exponent \(\nu \approx 0.94\) could partly be a consequence of the above mentioned unavoidable systematic uncertainties in current flux measurements near the bed.
More interestingly, it could also indicate that the splash for wind-blown sand grains is less efficient than for the plastic beads injected onto a quiescent bed in laboratory collision experiments~\cite{Beladjine2007,Laemmel2017}, from which one infers \(\nu\approx 1.4\).
The transport layer heights \(H\) obtained from the wind-tunnel data are in very good agreement with the obtained average value and the weak wind-dependence found in both our simulations and the analytical approach in Eq.~\eqref{eq:wind-depend-charact-trajectory} (right inset of Fig.~\ref{fig:flux-profiles}).
The field data, in contrast, yield an almost three times higher transport layer, which might be traced back to the non-equilibrium under-saturated transport conditions due to wind variations and/or due to moisture-induced stickiness of the used beach sand.
This is also supported by the overall transport rates reported in Ref.~\cite{Namikas2003} that are much smaller than those reported for wind-tunnel experiments.

\begin{figure}
  \centering
  \includegraphics{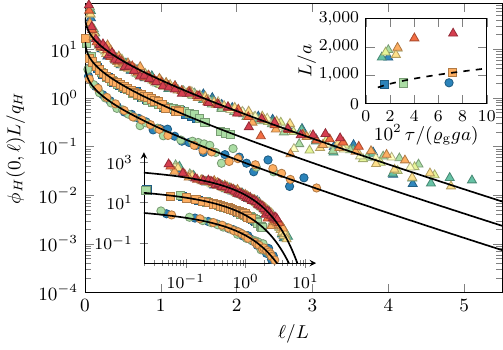}
  \caption{Theory and literature data for the vertical sand flux \(\phi_H(z=0,\ell)\) through the sand bed at a downwind distance \(\ell\) from the end of the sand bed (linear/logarithmic height axis; data extracted from independent sources vertically shifted, for better visibility).
    The simple analytical model with shape-invariant grain trajectories,  Eq.~\eqref{eq:vert-flux-profile}, is fitted to wind tunnel measurements by Ho \etal~\cite{Ho2014} (dots) and Rasmussen \etal~\cite{Rasmussen2015} (squares), and to field data by Namikas~\cite{Namikas2003} (triangles) for various wind strengths \(\tau\), using the value \(\nu=0.94\) obtained from Fig.~\ref{fig:flux-profiles}.
    The trajectory length \(L\) (the only free fit parameter) is compared to the theoretical expectation \(H/\epsilon(H) \approx 10 H\)~\cite{Laemmel2012} (dashed line) in the right inset.
    While laboratory data and theory agree well, field measurements consistently find longer (and higher) trajectories.}
  \label{fig:vert-flux}
\end{figure}

Similar conclusions can be drawn from the comparison between literature data  and the model prediction for the vertical flux \(\phi(z=0,\ell)\), in Fig.~\ref{fig:vert-flux}.
Again good agreement is obtained for \(\nu=0.94\) and using the characteristic hop length \(L\) as a free fit parameter.
The wind-tunnel data again agree well with the theoretical expectation \(L= H/\epsilon(H) \approx 10H\) based on Eq.~\eqref{eq:wind-depend-charact-trajectory}, whereas field data suggest higher values.

\begin{figure}
  \centering
  \includegraphics{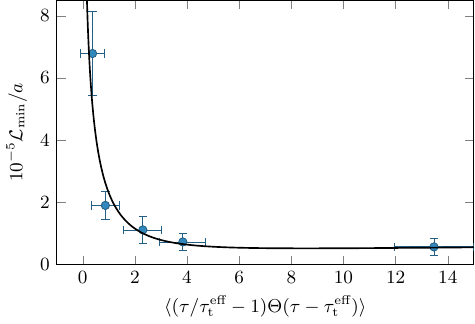}
  \caption{Field data for the minimum dune length \(\LL_\te{min}\) reported by Andreotti \etal~\cite{Andreotti2010} (symbols) compared to the dressed saturation length \(\avg{\phi\ell_\te{sat}^\te{dune}}/\avg{\phi}\) from Eqs.~\eqref{eq:saturation-length},~\eqref{eq:wind-depend-charact-trajectory} (line).
    The expected linear relation is recovered for Weibull-distributed intermittent wind-speed fluctuations of variance \(0.05\avg{u}^2\)~\cite{Pfeifer2012} and an effective threshold shear stress \(\tau_\te t^\te{eff} = 0.12 \tau_\te t\), suggesting a factor of proportionality of about \(35\) in Eq.~\eqref{eq:lsat-fluct-avg}, reasonably close to previous theoretical estimates~\cite{Kroy2005,Parteli2007a}.}
  \label{fig:min-dune-size}
\end{figure}

Besides such height-resolved characteristics for the stationary transport, our approach also gives access to its transient behavior, as shown in Sec.~\ref{sec:flux-transients}.
There, we argued that the minimum dune size measured in the field should be correlated with a ``dressed'' saturation length, averaged over some intermittent wind fluctuations.
This hypothesis is tested against field data for the minimum dune length \(\LL_\te{min}\)~\cite{Andreotti2010} in Fig.~\ref{fig:min-dune-size}.
The authors quantify the wind strength by the average \(\avg{(\tau/\tau_\te t^\te{eff} - 1)\Theta(\tau-\tau_\te t^\te{eff})}\).
Here \(\tau_\te t^\te{eff}\) is an effective transport threshold that is measured in the field under fluctuating wind conditions.
Intriguingly, this procedure yields a relatively small value \(\tau_\te t^\te{eff}\) compared to theoretical expectations and laboratory values for \(\tau_\te t\)~\cite{Greeley1985,Claudin2006,Shao2000a}, in line with our finding of a substantially renormalized threshold in Fig.~\ref{fig:salsim-saturation-length}.
Figure~\ref{fig:min-dune-size} moreover demonstrates good agreement of the field data with Eq.~\eqref{eq:lsat-fluct-avg} for Weibull-distributed wind speeds \(u\) with variance \(0.05\avg{u}^2\)~\cite{Pfeifer2012}, \(\tau_\te t^\te{eff} \approx 0.12 \tau_\te t\).
Even the ratio \(\LL_\te{min}\avg{\phi}/\avg{\phi \ell_\te{sat}^\te{dune}} \approx 35 \) between the minimum dune size and the dressed saturation length, which we used as a free fit parameter in the comparison, turns out to be reasonably close to (about \(1.5\) times larger than) the value predicted by numerical solutions of two- and three-dimensional versions of the minimal model for aeolian sand dunes~\cite{Kroy2005,Parteli2007a}.
This further supports our interpretation and underscores the importance of the distinction between bare and dressed mesoscale quantities in the analysis of field data.

For computational details and further theoretical and experimental support for an essentially constant value of \(\LL_\te{min}/\ell_\te{sat}^\te{dune}\) (rather than a constant \(\ell_\te{sat}^\te{dune}\)~\cite{Andreotti2010}), we refer the reader to App.~\ref{sec:minimum-dune-size}.

\section{Summary and conclusions}
\label{sec:summary}
We have developed an analytically tractable model for aeolian sand transport that resolves the whole distribution of grain trajectories.
Our analytical approach was heavily based on a recently proposed model for grain hopping on a granular bed that admits an analytical parametrization of the splash~\cite{Laemmel2017}, and guided and validated by coarse-grained computer simulations.
The core element of our model is the physically well grounded expression for the hop-height distribution in Eq.~\eqref{eq:drag-free-hopheight-distribution}, which is complemented by the ballistic approximations in Eqs.~\eqref{eq:hop-time}--\eqref{eq:concentration-profile-def} for the grain trajectories.
This combination allowed us to derive analytical predictions for various non-trivial mesoscale characteristics of the aeolian transport layer, among them the vertical and horizontal grain flux distributions and the saturation length, which we found to be in excellent agreement with an extensive compilation of independently generated field and wind-tunnel data using a consistent set of model parameters for the grain-scale physics.
Our approach thus correctly captures the splash and transport statistics and provides a canonical theoretical formalism for various height-resolved mesoscale properties.
It can be employed as a default in calculations when the actual profiles are not known, or as an alternative fit function (in place of the usual exponential) to extract parameters, such as the mean hop height, from experimental data.
Our detailed comparison of the analytical model and field and laboratory data revealed that it is necessary to distinguish so-called ``bare'' mesoscale relations (corresponding to precisely controlled ambient conditions) from their ``dressed'' counterparts that involve an average over fluctuations.
And it strongly suggests that our analytical mesoscale description provides a sound and versatile starting point for a precise and highly efficient modeling of a wealth of applications, from aeolian structure formation over dust emission to desertification.

\begin{acknowledgments}
  We thank Anne Meiwald for fruitful discussions during the early stages of this work, which was supported by a Grant from the German-Israeli Foundation for Scientific Research and Development (GIF).
  We also acknowledge the KITP in Santa Barbara and the MPI-PKS in Dresden for their hospitality during our visits, which were financially supported by the National Science Foundation under Grant No.\ NSF PHY-1125915, the MPI-PKS Visitors Program, and the German Academic Exchange Service (DAAD) through a Kurzstipendium for M.L.
  \end{acknowledgments}

\appendix
\section{\texorpdfstring{\(\varrho_H(z)\) and \(j_H(z)\)}{Height-dependent grain concentration and flux} for shape-invariant trajectories}
\label{sec:shape-inv-traj}
For shape-invariant trajectories, \ie, \(\epsilon(h)=\epsilon(H)\) independent of the hop height \(h\), Eqs.~\eqref{eq:drag-free-hopheight-distribution}--\eqref{eq:ver-velocity-vs-h} yield
\begin{equation}
  \label{eq:shape-invariant:concentration-profile}
  \begin{split}
    \frac{\varrho_H(z)}{\rho_H/H}
    = &\frac{\Gamma(1/2-\nu)}{2\Gamma(3/2-\nu)}  M( 1/2, 1/2+\nu, -z/H ) \\
    & + \frac{\sqrt{\pi} \Gamma(\nu-1/2)}{2 \Gamma(3/2-\nu) \Gamma(\nu)} (z/H)^{1/2-\nu} \\
    &\qquad  \cdot M(1-\nu,3/2-\nu,-z/H)
  \end{split}
\end{equation} and
\begin{equation}
    \label{eq:shape-invariant:flux-profile}
    \begin{split}
      \frac{j_H(z)}{q_H/H}
      = &\frac{\Gamma(1-\nu)}{2\Gamma(2-\nu)}  M( 1/2, \nu, -z/H ) \\
      &+ \frac{\sqrt{\pi} \Gamma(\nu-1)}{2 \Gamma(2-\nu) \Gamma(\nu-1/2)} (z/H)^{1-\nu} \\
      & \qquad \cdot M(3/2-\nu,2-\nu,-z/H)
  \end{split}
\end{equation}
for the grain mass concentration and flux, respectively.
Here,
\begin{equation}
  M(a,b,z) = \frac{\Gamma(b)}{ \Gamma(a) \Gamma(b-a)} \int_0^1 \!\!\! \df s\, \te e^{z s} s^{a-1}(1-s)^{b-a-1}
\end{equation}
denotes Kummer's confluent hypergeometric function and \(q_H = \Gamma(2-\nu)\Gamma(3/2-\nu)^{-1} \rho_H v_x(H)\) is the height-integrated grain flux.
The asymptotic forms in Eqs.~\eqref{eq:concentration-profile} and \eqref{eq:flux-height-profile} can be directly inferred from Eqs.~\eqref{eq:shape-invariant:concentration-profile} and \eqref{eq:shape-invariant:flux-profile}, respectively.
For \(z \gg H\), we may estimate the dependence of \(j_H(z) \propto [\epsilon(H)/\epsilon(z)] (z/H)^{1/2-\nu} \te e^{-z/H}\) on \(\epsilon(z)\) using Laplace's method to approximate the integral of the form
\begin{equation}
  \label{eq:profiles-laplace-approximation}
  \int_z^\infty \!\!\!\! \te d h \, f(h) \frac{ \te e^{-h/H}}{\sqrt{h-z}}
  \sim \sqrt{\pi H} \, f(z) \te e^{-z/H}
\end{equation}
after substituting \(h = z \cosh^2\theta\), expanding the exponent \((h/H)\cosh^2\theta\) around \(\theta=0\), and integrating over all \(\theta>0\).
 
\section{Median transport height and two-species prediction for \texorpdfstring{\(H\)}{H}}
\label{sec:two-spec}
The median height \(z_\te m\) of the transport layer is implicitly defined through
\begin{equation}
  \label{eq:two-spec:median-height}
  \begin{split}
    1/2 &= \int_0^{z_\te m} \hspace{-1.em} \df z \, j_H(z)/q_H \\
    &= 1/2 - M(-1/2,\nu-1,-z_\te m/H)/2  \\
    &\qquad + \frac{\sqrt{\pi} \Gamma(\nu-1)}{4\Gamma(\nu-1/2)\Gamma(3-\nu)} (z_\te m/H)^{2-\nu} \\
    & \qquad \quad \cdot M(3/2-\nu, 3-\nu, -z_\te m/H) \,,
  \end{split}
\end{equation}
where we inserted Eq.~\eqref{eq:shape-invariant:flux-profile} to evaluate the integral in the second line.
For \(\nu = 1.4\), as predicted from the splash model~\cite{Laemmel2017} underlying Eq.~\eqref{eq:drag-free-hopheight-distribution}, this relation can be solved numerically, which yields \(H \approx 5.8 z_\te m\), \ie, \(z_\te m \approx 0.17 H\).
The smaller value \(\nu=0.94\) that we used to fit the laboratory and field data in Figs.~\ref{fig:flux-profiles} and \ref{fig:vert-flux} corresponds to a larger median height of value \(z_\te m \approx 0.43 H\).

This relation can be used to make contact with the two-species model for aeolian sand transport proposed in Ref.~\cite{Laemmel2012}.
We therefor obtain the two-species prediction for \(z_\te m\) assuming that it lies above the reptation layer, where the (saltation) grain flux decays exponentially.
This yields
\begin{equation}
  \label{eq:two-spec:layer-height}
  H \approx 0.76 \ln \left[ 2/ \left( 1+q^\te{rep}/q^\te{sal} \right) \right] \left(v^\te{sal}\right)^2/(2g)\,,
\end{equation}
were \(q^\te{rep}\) and \(q^\te{sal}\) are the reptation and saltation contributions to the height-integrated grain flux, respectively.
The numerical prefactor that relates the saltation velocity \(v^\te{sal}\) (computed in Ref.~\cite{Laemmel2012}) to the vertical component \(v_z^\te{sal}(z=0)\) of the rebound velocity of a saltating grain (required here) is determined by fitting this relation to our simulation data for \(H\), shown in the inset of Fig.~\ref{fig:salsim-hopheightdistr}.
The so obtained value \(v_z^\te{sal}(z=0)/v^\te{sal}\approx 0.13\) of this effective restitution coefficient is in accord with what we expect from the splash model of Ref.~\cite{Laemmel2017}.

\section{Computer simulations}
\label{sec:simulations}
In our computer simulations, the trajectory of each grain is obtained by solving the equations of motion
\begin{equation}
  \label{eq:sim:eom}
  \dot{\vv v} = \frac{3 C_\te{D}}{4\sigma a} \abs{\vv u - \vv v} (\vv u - \vv v) + \vv g \,.
\end{equation}
Here, \(\vv v \equiv \vv v(t) = \dot {\vv r}(t)\) is the velocity of the grain at time \(t\), \(\vv r(t)\) its position, \(\vv u \equiv \vv u[\vv r(t),t]\) the wind velocity field, \(\vv g \equiv -g \vv e_z\) the gravitational acceleration, \(\sigma \equiv \varrho_\te g/\varrho_\te a\) the grain--air mass-density ratio, and \(a\) the grain diameter.
For the particle drag coefficient, we use the accurate expression~\cite{Jimenez2003}
\begin{equation}
  \label{eq:sim:drag-coefficient-jimenez2003}
  C_\te{D} = \frac{1}{3} \left[ A + \sqrt{A^2 +  16 B  \nu_\te{air}/(a \abs{\vv u-\vv v})} \right]^2
\end{equation}
(\(A \approx 0.95\) and \(B \approx 5.12\) for natural sand), where the viscous contributions to the drag are quantified by the Reynolds number \(a \abs{\vv u - \vv v}/\nu_\te{air}\), with the kinematic viscosity \(\nu_\te{air} \approx \SI{1.5}{\cdot 10^{-5}m^2 s^{-1}}\) for air.
To compute the grain trajectories with sufficient accuracy to  resolves the short hops of the low-energy ejecta, the equations of motion are solved using a standard Euler-forward integration scheme with a discretization time of \(0.1\sqrt{a/g}\) (corresponding to about \SI{0.45}{ms} for a grain diameter of \SI{200}{\micro m}).

Assuming a horizontally uniform stationary air flow in \(x\)-direction, the only non-vanishing component of the air velocity \(\vv u\) is given by the height-dependent wind speed \(u(z) \equiv \vv u \cdot \vv e_x \).
For constant pressure, the Reynolds-averaged Navier-Stokes equations reduce to the relation
\begin{equation}
  \label{eq:sim:RANS}
  \partial_z \tau_\te a(z) = F_x(z)
\end{equation}
between the \(xz\) component \(\tau_\te a\) of the Reynolds stress tensor for the air and the body force \(F_x(z)\) exerted by the grains on the air~\cite{Ungar1987}.
Without particles, \(F_x(z)=0\), this means that \(\tau_\te a = \tau\) is independent of \(z\) and given by the overall shear stress \(\tau\).
That this holds above the transport layer for \(z\to \infty\) provides us with the boundary condition for integrating Eq.~\eqref{eq:sim:RANS} for \(F_x(z) \neq 0\),
\begin{equation}
  \label{eq:sim:total-stress}
  \tau - \tau_\te a(z) = \int_z^\infty \!\!\!\! \!\! \df \tilde z \, F_x(\tilde z) \equiv \tau_{\te g, H}(z) \,,
\end{equation}
which defines the grain-borne shear stress \(\tau_{\te g, H}\)~\cite{Owen1964}.
In the simulations, the contribution of each grain to \(F_x\) is obtained as the negative \(x\)-component of the right-hand side of Eq.~\eqref{eq:sim:eom} times the grain mass \((\pi/6) \varrho_\te g a^3\).
The grain stress \(\tau_{\te g,H}\) is pre-averaged over a short time interval \(\sqrt{a/g}\) (ten time steps), in order to obtain smooth profiles.
With Eq.~\eqref{eq:sim:total-stress}, the Prandtl turbulence closure \(\kappa^2 \varrho_\te a z^2 [u'(z)]^2 = \tau_\te a(z)\) yields the height-dependent wind speed
\begin{equation}
  \label{eq:sim:prandtl-closure}
  u(z) = \frac{1}{\kappa \sqrt{\varrho_\te a}} \int_{z_0}^z \df \zeta \, \frac{ \sqrt{\tau - \tau_{\te g, H}(\zeta)}}{\zeta}  \,.
\end{equation}
The roughness height \(z_0\), where the wind speed nominally vanishes, is set to \(a/10\) by convention.

Equations \eqref{eq:sim:eom}--\eqref{eq:sim:prandtl-closure} have to be complemented by a set of boundary conditions describing the sand bed collisions.
They are taken from an extended version of the coarse-grained splash model proposed in Ref.~\cite{Laemmel2017} that accounts for wind-induced lift forces acting on the bed grains, as briefly outlined in App.~\ref{sec:splash-model}.
To keep the computation simple, we neglect midair collisions.
They are most frequent in the dense reptation layer close to the bed~\cite{Huang2007,Duran2014} and therefore probably well captured by appropriate effective bed parameters, except under extreme wind conditions~\cite{Carneiro2013}.

\begin{figure}
  \centering
  \includegraphics{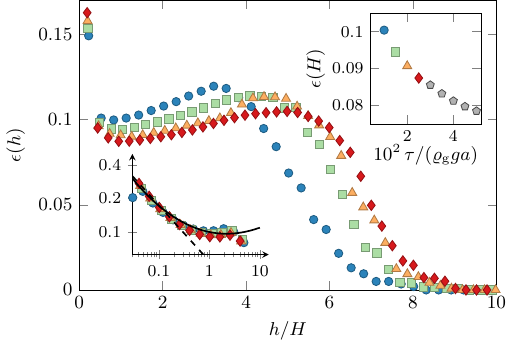}
  \caption{Trajectory aspect ratio \(\epsilon \equiv h/l\) for various wind strengths, as obtained from our simulations.
    For most mesoscale measures, its dependence on the hop height \(h\) is negligible, as is the weak dependence on the wind shear stress \(\tau\), as illustrated in the right inset for the aspect ratio \(\epsilon(H)\) of the characteristic trajectory.
    However, the height-resolved grain stress, Eq.~\eqref{eq:grain-stress}, crucially depends on the power-law decay for \(h \ll H\), which is better resolved on the logarithmic scale in the left inset, solid and dashed lines representing Eq.~\eqref{eq:splash:ejecta-aspect-ratio} (scaled by an empirical prefactor \(0.65\)) and the power-law \(\epsilon \propto h^{-1/2}\), respectively.}
  \label{fig:salsim-aspectratio}
\end{figure}

\section{Splash model}
\label{sec:splash-model}
For convenience, we now outline the key properties of the splash model of Ref.~\cite{Laemmel2017} used to quantify the sand bed collisions in the numerical simulations (App.~\ref{sec:simulations}) and in our analytical derivation of  Eq.~\eqref{eq:drag-free-hopheight-distribution} for the hop-height distribution.
Its original version was developed assuming vacuum conditions.
For our present purpose, we extend it to effectively account for the drag and lift forces exerted on the bed grains by the wind.

The rebound of a hopping grain from a rough immobile bed is quantified in terms of the mean total restitution coefficient \(\overline e\) and rebound angle \(\theta'\) for a given impact angle \(\theta\).
Averaging over all possible collision geometries gives
\begin{equation}
  \label{eq:splash:total-restitution}
  \overline e \equiv \abs{\vv v'}/\abs{\vv v} = \beta_\te s - (1-\alpha_\te s^2/\beta_\te s^2) \beta_\te s \theta/2 \,,
\end{equation}
where \(\alpha_\te s \approx 0.20\) and \(\beta_\te s \approx 0.63\) were calibrated~\cite{Laemmel2017} with collision experiments by Beladjine \etal~\cite{Beladjine2007}.
The rebound angle \(\theta'\) is randomly drawn from the distribution
\begin{equation}
  \label{eq:splash:rebound-angle-distribution}
  P(\theta' \vert \theta)
  = (2/\gamma_\te s)(1 + \theta'/ \theta) \ln \!\left[\gamma_\te s \theta/(\theta + \theta')^2\right]
\end{equation}
defined in the interval \(0 < \theta' + \theta < \sqrt{\gamma_\te s \theta} \) and set to zero outside.
We multiplied the parameter \(\gamma_\te s = (9/2) (1+\alpha_\te s/\beta_\te s)^2\) by an order-unity factor \(3\sqrt 3/4 \approx 1.3\) for harmonization with the approach underlying Eq.~\eqref{eq:splash:total-restitution}.
Note that Eq.~\eqref{eq:splash:rebound-angle-distribution} can yield negative rebound angles, corresponding to a grain that gets trapped in the bed.
Rebounding grains with \(\abs{\vv v'} \sin \theta' \geq \sqrt 2\) fail to leap over the downwind bed neighbor grain and are thus not propagated further.
To avoid discretization artifacts at the ground, which become apparent in height-resolved quantities like the grain concentration or the grain shear stress, we distribute the initial height of a grain ricocheted off the bed uniformly between \(z=0\) and \(z = \delta t \abs{\vv v'} \sin \theta'\), where \(\delta t\) is the duration of one time step in the simulations.

To mimic the statistics of the ejecta close to the ground, we set the ejection angle to \SI{90}{\degree} and draw the kinetic energy of the ejected grains form the log-normal distribution~\cite{Ho2012}
\begin{equation}
  \label{eq:splash:fragmentation-energy-distribution}
  P(E_\te{ej}' \vert E) = \frac{1}{\sqrt{2\pi} \sigma E_\te{ej}'} \exp \! \left[-\frac{(\ln E_\te{ej}'-\mu)^2}{2\sigma^2} \right]
\end{equation}
with the energy \(E=m\vv v^2/2\) of the impacting grain of mass \(m = (\pi/6)\varrho_\te g g a^3\) and \(\sigma = \sqrt \lambda \ln 2\), \(\mu = \ln [(1-\overline e^2)E] - \lambda \ln 2\), \(\lambda = 2 \ln\!\left[\left(1-\overline e^2\right)E/E_a\right]\), and \(E_a = mga\).
The number 
\begin{equation}
  \label{eq:splash:number-ejected-particles}
  N_2' = 0.06 [(1-\overline e^2) E/E_a ]^\varkappa \int_{E_\te{eff}}^\infty
 \df E_2' \, P(E_2'|E) \\
\end{equation}
of ejected bed grains again follows from the same energy balance approach underlying Eq.~\eqref{eq:splash:fragmentation-energy-distribution}, with the value \(\varkappa = (2-\ln 2) \ln 2 \approx 0.9\) of the exponent close to one.

In Eq.~\eqref{eq:splash:number-ejected-particles}, we extended the original splash model of Ref.~\cite{Laemmel2017} to account for drag and lift forces exerted on the bed grains.
We therefor introduced an effective minimum energy \(E_\te{eff}<E_a\) required for a bed grain to be mobilized during the collision.
Rewriting it in terms of an effective mass \(m_\te{eff} \equiv E_\te{eff}/(ga)\) of a bed grain exposed to the turbulent shear flow, we estimate its value following Bagnold's classic prediction of the shear stress value at the threshold of grain movement~\cite{Bagnold1941}.
Balancing the net torque due to the horizontal drag force \(F_\te D\) and vertical gravitational force \(mg\) acting on a bed grain of mass \(m\) with the purely gravity-induced torque of a grain of same size but different mass \(m_\te{eff}\), one obtains \(\sin(\beta_\te s) F_\te{D} - \cos( \beta_\te s) m = -\cos(\beta_\te s) m_\te{eff}\), for an angle \(\beta_\te s\) between the horizontal and the axis crossing the point of support and the center of the lifted bed grain.
This yields
\begin{equation}
  \label{eq:splash:drag-effective-mass}
  E_\te{eff}/E_a = m_\te{eff}/m
  = 1 - \tau_\te a(z=0)/\tau_\te{ta} \,.
\end{equation}
Here, we expressed the drag force \(F_\te{D} \propto \pi (a/2)^2 \tau_\te a(z=0)\) in terms of the air shear stress at the ground and identified its threshold value \(\tau_\te{ta} \propto (2/3)b/[\tan(\beta_\te s)]\), above which bed grains are mobilized by the flowing air (corresponding to a vanishing effective weight, \(m_\te{eff}=0\)).
We set \(\tau_\te{at} = 1.2\cdot 10^{-2} \varrho_\te g g a\), yielding a minimum shear stress \(\tau_\te{t} \approx 0.9 \cdot 10^{-2} \varrho_\te g g a\) required to maintain transport (once initialized), which is in agreement with typical values for the ratio \(\tau_\te{at}/\tau_\te t \approx 0.8\) observed in wind tunnel experiments~\cite{Bagnold1941,Carneiro2015}.

\section{Splash predictions for \texorpdfstring{\(\nu\), \(H\),  \(\epsilon\),  \(\alpha\)}{nu, H, epsilon, alpha}}
\label{sec:splash-predictions}
The quantitative description of the bed collisions outlined in App.~\ref{sec:splash-model} allows us to estimate typical values for the power-law exponent \(\nu\), the trajectory aspect ratio \(\epsilon(H)\), and the effective restitution coefficient \(\alpha\) introduced in Sec.~\ref{sec:reduced-model} as model parameters.

Relating the ejection energy \(E_\te{ej}' = m g h\) to the hop height, Eq.~\eqref{eq:splash:fragmentation-energy-distribution} provides us with the hop-height distribution per impacting grain of energy \(E\).
As \(E \gg mga\), \ie, \(\mu \gg \ln (mga)\), it is expanded to \(\propto h^{-\nu}\), with \(\nu\) given in Eq.~\eqref{eq:hopheight-nu}.
Multiplying it with the hop time \(T(h)\), Eq.~\eqref{eq:hop-time}, it yields the hop-height distribution \(P_H(h\ll H) \propto h^{1/2-\nu}\) per time step.

Next, we estimate the characteristic height \(H\) of the transport layer at low-wind conditions close to the transport threshold employing the splash-model prediction for collisions between saltating grains and the bed grains.
We start with the relation
\begin{equation}
  \label{eq:splash:hop-height-rebound-prob}
  H = - \overline{E_\te{ej}}(E)/(mg \log P_\te{reb})
\end{equation}
between \(H\) and the rebound probability \(P_\te{reb}\) derived in the text below Eq.~\eqref{eq:drag-free-hopheight-distribution}.
Here, we expressed the mean ejection height \(\overline{h_\te{ej}}(E) = \overline{E_\te{ej}}(E)/(mg)\) for a given impact energy \(E\) in terms of the mean ejection energy \(\overline{E_\te{ej}}(E)\).
The latter is obtained from the distribution in Eq.~\eqref{eq:splash:fragmentation-energy-distribution} as
\begin{equation}
  \label{eq:splash:mean-ejection-energy}
  \begin{split}
    \overline{E_\te{ej}}(E)
    &= \frac{\int \df E_\te {ej}' P(E_\te{ej}' \vert E) \Theta(E_\te{ej}'-E_a) E_\te{ej}'}{\int \df E_\te {ej}' P(E_\te{ej}' \vert E) \Theta(E_\te{ej}'-E_a)} \,, \\
    &= \frac{\erfc[(\ln E_a-\mu-\sigma^2)/(\sqrt 2 \sigma)]}{\erfc[(\ln E_a-\mu)/(\sqrt 2 \sigma)]} \te e^{\mu+\sigma^2/2} \,,
  \end{split}
\end{equation}
where the integrals extend over all energies \(E_\te {ej}'>E_a\) above the minimum ejection energy \(E_a = mga\).
As the impact energy \(E\) is determined by the difference \(u(H) - v_\infty\) between the wind speed at height \(H\) and the turbulent settling velocity \(v_\infty \approx 27 \sqrt{ga}\), we may approximate it as
\begin{equation}
  \label{eq:splash:impact-energy}
  E \approx (m/2)[u(H)-v_\infty]^2 \,.
\end{equation}
Making use of the low-transport conditions assumed here, the wind-speed profile in Eq.~\eqref{eq:wind-height-profile} simplifies to \(u(z) = (u_{\ast\te t}/\kappa) \ln(z/z_0)\), with \(u_{\ast\te t}^2/(ga) \approx 0.01 \sigma \approx 20\), \(\kappa = 0.4\) and \(z_0 \approx a/10\).
We can now insert this relation, together with Eqs.~\eqref{eq:splash:mean-ejection-energy}, \eqref{eq:splash:impact-energy} and the rebound probability \(P_\te{reb} \approx P_\te{reb}(h \gg a) = \int_0^\infty \df \theta' P(\theta'|\theta)\) as predicted by Eq.~\eqref{eq:splash:rebound-angle-distribution}, into Eq.~\eqref{eq:splash:hop-height-rebound-prob}, which yields \(H(\theta)\) as a monotonically decreasing function of the impact angle \(\theta\).
For reasonable values \(\theta = \SI{5}{\degree}\dots\SI{20}{\degree}\), it varies within the range \(H/a \approx 190 \dots 54\).
To determine its actual value, we use the relation \(\epsilon(H) \approx \tan(\theta)/4\) between trajectory aspect ratio and impact angle, as inferred from Eq.~\eqref{eq:hor-velocity-vs-h}, and combine it with
\begin{equation}
  \label{eq:splash:trajecotry-aspect-ratio}
  \epsilon(H) = H/L \approx H/\{T(H)[u(H)-v_\infty]\} \,,
\end{equation}
where we approximated the hop length with the flight time \(T(H)\), given in Eq.~\eqref{eq:hop-time}, times the typical grain speed.
Putting all together, we obtain a transcendental equation for the impact angle \(\theta\):
\begin{equation}
  \label{eq:splash:theta-condition}
  \cot(\theta) 4H(\theta)/T[H(\theta)] \approx u[H(\theta)]-v_\infty \,.
\end{equation}
With the parameters listed above, it is solved by \(\theta \approx \SI{14.5}{\degree}\), for which Eqs.~\eqref{eq:char-hopheight} and \eqref{eq:epsH} yield \(H = H(\theta) \approx 72 a\) and \(\epsilon(H)\approx 0.1\), respectively.

While trajectories of height on the order of \(H\) are crucially influenced by the impactor--bed rebound, the shape of small trajectories of height \(h \ll H\) is mainly dictated by the characteristics of the ejecta created during these bed collisions.
To estimate the height dependence \(\epsilon(h)\) of their aspect ratio, we simplify the equations of motion, Eq.~\eqref{eq:sim:eom}, as follows.
Since \(v_z \ll u\), vertical and horizontal motion decouple, and thus 
\begin{equation}
  \label{eq:splash:ejecta-eom}
  \dot v_x \approx L_\te{D}^{-1} [u(h)-v_x]^2 \,,
\end{equation}
with the constant drag length \(L_\te{D} = 4\sigma a/(3 C_\te{D}^\infty)\), where \(C_\te{D}^\infty \approx 1.2\) is the strong-turbulence limit of the drag coefficient computed from Eq.~\eqref{eq:sim:drag-coefficient-jimenez2003} for vanishing air viscosity \(\nu_\te{air}=0\), \ie, \(\te{Re}_\te g \to \infty\).
We further replaced the height-dependent wind speed \(u(z)\) in the drag relation by the constant value \(u(h)\) taken at the height of the trajectory.
It is computed using the above relation for threshold conditions (which also holds for \(\tau>\tau_\te t\), as \(u(z<H)\) exhibits the universal \(\tau\)-independent shape obtained in Eq.~\eqref{eq:wind-height-profile}).
For the initial condition \(v_x(t=0) = 0\) and the total flight time \(T(h)\), the time integral of the solution \(v_x(t) = u(h)^2 t/[L_\te D + u(h) t]\) of Eq.~\eqref{eq:splash:ejecta-eom} readily yields the hop length
\begin{equation}
  \label{eq:splash:ejecta-aspect-ratio}
  h/\epsilon(h)
  \approx T(h) u(h) - L_\te D \ln[1 + T(h) u(h)/L_\te D] \\
\end{equation}
for \(h \ll H\).
While the overall value of \(\epsilon(h)\) obtained from this estimate is found to be a bit too large (by a factor of \(1/0.65\approx 1.5\)), its qualitative form is indeed in very good agreement with our simulations (even for intermediate heights \(h \simeq H\)), as illustrated in Fig.~\ref{fig:salsim-aspectratio}.
Inserting Eq.~\eqref{eq:hop-time} for \(T(h)\) together with the scaling relations \(u(h) \sim \sqrt{\sigma g a} \ln(h/z_0)\) and \(L_\te D \sim \sigma a\), and expanding the right hand side of Eq.~\eqref{eq:splash:ejecta-aspect-ratio} for \(h \ll \sigma a\) (equivalent to \(T(h) u(h) \ll L_\te D\)), it becomes \(\epsilon(h) \propto \ln(h/z_0)^{-2}\).
For \(h \gg z_0\), the \(h\)-dependence is (locally) very close to a power law with an exponent of value \(-2/\ln(h/z_0)\) that varies only weakly as \(h\) gets larger.
Substituting the typical ejection height \(\avg{h}_H \approx 7 a\) obtained in Eq.~\eqref{eq:mean-hop-height} (or \(\overline{h_\te{ej}}_H \approx 10 a\) from \eqref{eq:splash:mean-ejection-energy}) for \(h\), this exponent takes the value \(0.47 \approx 1/2\), which gives rise to the first line of Eq.~\eqref{eq:aspect-ratio-regimes}.
The applicability of this power-law approximation for \(h \ll H\) is supported by the comparison with our simulations in Fig.~\ref{fig:salsim-aspectratio}.

We eventually derive a splash-model prediction for the effective restitution coefficient \(\alpha\), introduced in Eq.~\eqref{eq:ver-velocity-increase-vs-h}.
Identifying \(v_x(h)\) with the average of the horizontal impact and rebound velocities, it is related to the total restitution \(\overline e\) and the impact and rebound angle \(\theta\), \(\overline{\theta'}\) via
\begin{equation}
  \label{eq:splash:alpha}
  \alpha = 2(\cos \theta - \overline e \cos \overline{\theta'})/(\cos \theta + \overline e \cos \overline{\theta'}) \,.
\end{equation}
With Eqs.~\eqref{eq:splash:total-restitution}, \eqref{eq:splash:rebound-angle-distribution}, \(\alpha\) takes values between \(0.45\) (for \(\theta=0\)) and \(0.64\) (at \(\theta \approx \SI{28}{\degree}\)).
For the typical impact angle \(\theta \approx \SI{13}{\degree}\) obtained from Eq.~\eqref{eq:splash:theta-condition}, we get \(\alpha \approx 0.6\).

\section{Minimum dune size}
\label{sec:minimum-dune-size}
We provide a theoretical estimate for the minimum dune length \(\LL_\te{min}\), for which literature field-data are shown in Fig.~\ref{fig:salsim-saturation-length}.
To this end, we employ a slightly improved version of the linear stability analysis proposed in Refs.~\cite{Andreotti2002a,Fourriere2010}.
It identifies \(\LL_\te{min}\) with the wavelength of the (initially) fastest growing mode of a weakly perturbed flat sand bed.
The starting point is the out-of-phase response of shear stress and sand transport to weak perturbations of a flat bed~\cite{Kroy2002,Kroy2002a}.
In Ref.~\cite{Fourriere2010}, Andreotti and coworkers used the generic formula \(q_\te{sat} \propto \tau^\chi(\tau-\tau_\te t)\) for the saturated flux, which is \emph{indirectly} slope-dependent via the variation of the local shear stress \(\tau\) and the threshold value \(\tau_\te t\).
However, they neglected the \emph{direct} slope dependence, which obtains even for fixed wind parameters, and should be accounted for, to first order in \(h'\), according to
\begin{equation}
  q_\te{sat} \propto \tau^\chi(\tau-\tau_\te t)(1-h' \cot \theta_\te f) \,,
  \label{eq:qsat-slopedependent}
\end{equation}
where \(\theta_\te f\) is the friction angle.
This direct slope dependence of \(q_\te{sat}\) was first predicted by
Bagnold and later validated by Iversen and Rasmussen
\cite{Iversen1999}, who obtained \(\theta_\te f \approx 40^\circ\) for
typical dune sand (with grain diameter \(d \geq \SI{170}{\micro m}\)).
Equation \eqref{eq:qsat-slopedependent} can also be derived from the continuum saltation model by Sauermann \etal~\cite{Sauermann2001} if we extend its kinematics to sloped beds.
Accounting for the gravitational force \(- g h'\) in the momentum balance, Eq.~(33) in Ref.~\cite{Sauermann2001}, we obtain
\begin{equation}
  q_\te{sat} \propto (v_0/g)(\tau-\tau_\te t)(1 - h' \alpha v_\infty/v_0) \,.
  \label{eq:sauermann-slopedep-flux}
\end{equation}
Here, the turbulent settling velocity \(v_\infty\) accounts for the bed friction and is equal to the difference between the wind speed and the steady-state transport velocity \(v_0\) over the unperturbed bed.
For moderate wind strengths, \ie, \(\tau \approx \tau_\te t\), \(v_0\) is independent of \(\tau\), corresponding to \(\chi=0\), and the coefficient \(\alpha v_\infty/v_0\) takes values on the order of 0.5 for typical dune sand, which corresponds to \(\theta_\te f \approx 60^\circ\) in Eq.~\eqref{eq:qsat-slopedependent}, in reasonable agreement with the phenomenological law.

For the stability analysis, we follow Fourri\'ere \etal~\cite{Fourriere2010} and expand the perturbation \(\delta q_\te{sat} = q_\te{sat} - q_{\te{sat} 0}\) of the saturated flux from its flat bed value \(q_{\te{sat} 0}\) to first order in the perturbation \(\delta h\) of the bed profile.
From Eq.~\eqref{eq:qsat-slopedependent}, we obtain its Fourier transform \(\delta \hat q_\te{sat} = \int \!
\te d x \, \delta q_\te{sat} \, \te e^{-i k x} \propto \tau_0^\chi (A_q + i B_q) k \, \delta \hat h \) with
\begin{equation}
  \label{eq:flux-perturbation-ab}
  \begin{split}
    A_q &= A_\tau + \chi A_\tau(1- \tau_{\te t 0}/\tau_0) \quad
    \te{and} \\
    B_q &= B_\tau + \chi B_\tau(1- \tau_{\te t 0}/\tau_0) - \cot \theta_\te f \,,
  \end{split}
\end{equation}
where we used the shear stress perturbation \(\delta \hat \tau/\tau_0 = (A_\tau + i B_\tau) k\) (for \(k \geq 0\)) and the slope-dependent threshold \(\delta \hat \tau_\te t = i k \delta \hat h \cot \theta_\te f\).
Following the convention, we denote the unperturbed shear stress over a flat ground far away from the obstacle by \(\tau_0\) (the subscript was suppressed in the main text).
Only the \(\tau_0\)-independent term \(\cot \theta_\te f\) in the expression for \(b\) differs from the result \((\tau_\te t/\tau_0) \cot \theta_\te f\) by Fourri\'ere \etal, who neglected the direct slope dependence of \(q_\te{sat}\).
While this may seem to be a minor change of the original argument, it yields a qualitatively different result.
Combining the linearized differential equation for the flux evolution with mass conservation \(\partial_t h \propto - \partial_x q\) for the sand bed profile \(h\), the ratio \(\LL_\te{min}/\ell_\te{sat}^\te{dune}\) of the wavelength of the fastest growing mode to the saturation length follows as a function of the ratio \(A_q/B_q\).
For \(\LL_\te{min}/\ell_\te{sat}^\te{dune} \gg 1\), it scales as
\begin{equation}
  \label{eq:fastest-growing-mode-approx}
  \LL_\te{min}/\ell_\te{sat}^\te{dune} \sim 3\pi A_q/B_q \,.
\end{equation}
Substituting \(A_q\) and \(B_q\) from Eq.~\eqref{eq:flux-perturbation-ab}, we see that \(\LL_\te{min}/\ell_\te{sat}^\te{dune}\) decreases with increasing \(\tau_0\) as long as \(\chi>0\).
This is qualitatively similar to what was obtained by Fourri\'ere \etal\ neglecting the experimentally well established and theoretically derived rightmost factor in Eq.~\eqref{eq:qsat-slopedependent}.
However, for the theoretically expected and generally accepted value \(\chi=0\), the ratio \(A_q/B_q = A_\tau/(B_\tau-\cot \theta_\te f)\) becomes independent of \(\tau_0\), and so does \(\LL_\te{min}/\ell_\te{sat}^\te{dune}\).
Evidently, the direct slope dependence of the flux, represented by the friction angle \(\theta_\te f\), can be understood as an effective renormalization of the symmetry-breaking part of the driving wind field perturbation, quantified by \(B_\tau\).
This correction crucially affects the absolute value of \(A_q/B_q\), which diverges at \(B_\tau \to \cot \theta_\te f \approx 1.2\).
The numerical values of the coefficients \(A_\tau\) and \(B_\tau\) can be estimated as functions of the dimensionless hydrodynamic bed roughness \(\eta_0\).
Using for simplicity the analytical dependencies \(A_\tau(\eta_0)\) and \(B_\tau(\eta_0)\) calculated by Hunt and coworkers~\cite{Hunt1988} (see Fig.~2 of Ref.~\cite{Kroy2002a}), the divergence of \(A_q/B_q\) is expected near \(\eta_0\approx 1.7 \times 10^{-5}\).
We can match the fit result \(\LL_\te{min}/\ell_\te{sat}^\te{dune} \approx 35\) obtained from Fig.~\ref{fig:salsim-saturation-length} for \(\eta_0 \approx 4.0 \cdot 10^{-3}\), which lies in the estimated range \(10^{-4}\) to \(10^{-2}\)~\cite{Fourriere2010} for \(\eta_0\) for typical sand dunes (corresponding to \(\LL_\te{min}/\ell_\te{sat}^\te{dune} \approx 30 \dots 140\)).

In summary, we showed that an improved version of the linear stability analysis proposed by Fourri\'ere \etal\ yields a \(\tau_0\)-independent ratio \(\LL_\te{min}/\ell_\te{sat}^\te{dune}\) if the experimentally and theoretically well established form of the wind-strength dependent sand flux given by Eq.~\eqref{eq:qsat-slopedependent} with \(\chi=0\) is employed.
Together with the experimentally observed wind dependence of the minimum dune size \(\LL_\te{min}\) (see Fig.~\ref{fig:salsim-saturation-length}), this implicates that the saturation length \(\ell_\te{sat}^\te{dune} \propto \LL_\te{min}\) must strongly decrease with increasing wind strength.
It is argued in the main text that this can be rationalized by a strong renormalization of the effective saturation length by intermittent turbulent wind-strength fluctuations near the threshold rather than by a rapid decay of \(\LL_\te{min}/\ell_\te{sat}^\te{dune}\) with increasing \(\tau_0\) and a constant saturation length, as previously proposed~\cite{Fourriere2010,Andreotti2010}.

\bibliographystyle{apsrev4-1}
\bibliography{sand}
\end{document}